\documentclass[preprint,aps,twocolumm,showpacs]{revtex4}
\usepackage{amsfonts}
\usepackage{amsmath}
\usepackage{amssymb}
\usepackage{graphicx}

\begin{document}
\title{Resonances on deformed thick branes}
\author{W. T. Cruz$^1$, A. R. Gomes$^2$, C. A. S. Almeida $^3$}
\affiliation{$^1$
 Instituto Federal do Cear\'a, Campus Juazeiro do Norte, 63040-000 Juazeiro do Norte - Cear\'{a} - Brazil\\ E-mail: {wilami@fisica.ufc.br}\\
 $^2$Instituto Federal do Maranh\~ao, Campus Monte Castelo, \\ S\~ao Lu\'is - Maranh\~ao - Brazil\\
E-mail: {argomes@pq.cnpq.br} \\
$^3$Departamento de
F\'{\i}sica - Universidade Federal do Cear\'{a} \\ C.P. 6030, 60455-760 Fortaleza - Cear\'
{a}-Brazil\\E-mail: {carlos@fisica.ufc.br}}

\begin{abstract}
In  this work we investigate the issue of gravity and fermion localization and resonances in $(4,1)$-branes constructed with one scalar field coupled with gravity in deformed models. Such models give solutions for the scalar field that is the usual kink solution in the extra dimension for a parameter $p=1$ and deformations with a two-kink profile for odd $p>1$. Gravity is localized and resonant modes are found for small values of $p$. The coupling between the scalar field and spinors is a necessary condition for fermions to be localized on such branes. After performing a chiral decomposition of the five-dimensional spinor we found resonances with both chiralities for all odd $p$'s. The correspondence between the spectra for left and right chirality is guaranteed and Dirac fermions are realized on the brane. The increasing of $p$ characterizes the formation of branes with internal structure that turns the gravitational interaction more effective for fermions aside the brane, increasing their lifetime. The influence of the internal structure of the branes and the presence of resonances for gravity and fermionic modes is addressed.
\end{abstract}

\maketitle

\section{Introduction}
Brane structures where initially introduced as domain walls embedded in extra dimensions \cite{cs,akama,rub,vis}. In a sense thick branes \cite{de,csaba,gremm,fase,bc} are a natural construction since the solutions can be dynamically found. More recently, interesting models have been proposed for such dynamical branes with rich structures, constructed with one \cite{adalto} or more \cite{brane,es,dutra} scalar fields (see also the review given by Ref. \cite{dzhun}). The issue of localization of several fields and resonances in such branes is an interesting subject, as their investigation can guide us to which kind of brane structure is more acceptable phenomenologically. In this way, models which extend the Randall-Sundrum type II scenario \cite{rs2} were constructed and considered under the aspect of gravity localization \cite{man,mel,bgl,bh,bhr}.

Reviewing works about fermions in extra dimension models, we observed that in type I Randall-Sundrum (RS) scenario \cite{rs1}, spin $1/2$ and spin $3/2$ fermions, are located only in the negative tension brane. However, they do not satisfy the localization conditions in the type II RS model. In these two cited models, the localization of zero mode chiral fermions is only obtained due to the introduction of a generalized Yukawa coupling. This is analogous to what happens to domain walls in the absence of gravity \cite{jackiw}. In the brane case the coupling is introduced by an interaction of fermions with a bulk scalar field. The scalar field solution can be kink like, for instance. Thus, the results with the RS model indicate that the thick brane scenarios as the most suitable for locating fermions. Thick brane models can be obtained from scalar fields with kink-like solutions. Therefore, the scalar field that couples with fermions in the Yukawa coupling will be the scalar field which the brane is made of.

Quite recently, Liu et al. \cite{Liu1} have analyzed the issue of fermion localization on a pure geometrical thick brane. Since then, the same group of authors have published a series of papers on fermion localization on several types of thick branes \cite{Liu2,Liu3,Liu4,Liu5}. Also, some of the authors of the present work were the first to consider fermionic resonances in branes with internal structure \cite{ca}.

As is well known, the kind of structure of the considered brane is very
important and will produce implications concerning the methods of
field localization. In the seminal works of Bazeia and collaborators \cite{deformed,aplications} a class
of topological defect solutions was constructed starting from a
specific deformation of the $\phi^4$ potential. These new solutions
may be used to mimic new brane-worlds containing internal structures
\cite{aplications}. Such internal structures have implications in the
density of matter-energy along the extra dimensions \cite{brane} and
this produces a space-time background whose curvature has a
splitting, as we will show in this work, if compared to the usual models. Some
characteristics of such model were considered in phase transitions
in warped geometries \cite{fase}.

Considering a brane world scenario like a 4D domain wall immersed in a 5D space-time, we were able to find a localized left chiral zero mode. Therefore, if we consider a deformed brane, we obtain new results related to the localization of spin $1/2$ fields. Also, such class of models where already studied with respect to gravity localization \cite{adalto}, where it was found that zero-modes for KK gravitons exist, the solutions are stable and tachyons are forbidden.

The main goal here is to study the behavior of fermions and gravity in a membrane with internal structure generated by a deformation procedure. We modify the so-called two-kink solutions that can be obtained after a deformation procedure of a potential from a scalar field \cite{deformed}. We analyze fermion localization on such branes, considering such deformations that suggest the existence of an internal structure. As we will see, the deformations will be very important for localization and normalization of fermionic fields. Using a well known resonance detecting method \cite{Liu4,ca,wilami1}, we analyze the massive modes arising from the dimensional reduction functions. Modes with large amplitudes near the brane can reveal us details about the coupling of modes with the brane.

 This paper is divided as follows. In the next section we review the deformed brane model with one scalar field, where we analyze the bulk scalar curvature, relating the results to the the brane internal structure. Sec. III is devoted to the analysis of gravity localization and resonances with KK gravitons. Each particular brane solution is considered as a fixed background that are not perturbed significantly by the presence of fermions. With this approximation, in Sec. IV we perform a chiral decomposition and study the presence of zero modes with left chirality. Massive modes and their normalization are subject of investigation in Sec. V. Fermionic resonances and the realization of Dirac fermions on the brane are considered in Sec. VI. The main conclusions are presented in Sec. VII.

\section{Brane setup}

We start with the action describing one scalar field minimally coupled with gravity in five dimensions
\begin{equation}
S=\int d^{5}x \sqrt{-G}[2M^{3}R-\frac{1}{2}(\partial\phi)^{2}-V(\phi)],
\end{equation}
where $M$ is the Planck constant in $D=5$ dimensions and $R$ is the  scalar curvature. For some classes the potential $V(\phi)$, it is possible to obtain kink solutions for the field $\phi$ depending only on the extra dimension. As an Ansatz for the metric we consider an extension for the Randall-Sundrum metric, where the bulk spacetime is asymptotically $AdS_5$, with a Minkowski brane,
\begin{equation}\label{metrica}
ds^{2}=e^{2A(y)}\eta_{\mu\nu}dx^{\mu}dx^{\nu}+dy^{2}
\end{equation}
The warp factor depends on the metric function $A(y)$, where $y$ is the extra dimension. The tensor $\eta_{\mu\nu}$ is the Minkowski metric and the indices
$\mu$ and $\nu$ vary from 0 to 3. For this background we find the following equations of motion:
\begin{equation}
\phi^{\prime\prime}+4A^\prime\phi^\prime=\frac{dV}{d\phi},
\end{equation}
\begin{equation}\label{mov1}
\frac{1}{2}(\phi^{\prime})^{2}-V(\phi)=24M^{3}(A^{\prime})^{2},
\end{equation}
and
\begin{equation}\label{mov2}
\frac{1}{2}(\phi^{\prime})^{2}+V(\phi)=-12M^{3}A^{\prime\prime}-24M^{3}(A^{%
\prime})^{2}.
\end{equation}
Here prime means derivative with respect to the extra dimension.

In the presence of gravity, defining the potential as
\begin{equation}
V_p(\phi)=\frac{1}{2}\left(\frac{dW}{d\phi}\right)^2-\frac{8M^3}{3}W^2,
\end{equation}
it is possible to find first-order equations
\begin{equation}
\label{1order}
\phi'=\frac{\partial W}{\partial \phi},
\end{equation}
\begin{equation}
\label{WA}
W=-3A'(y),
\end{equation}
 whose solutions are also solutions from the equations of motion. Here $W(\phi)$ is the superpotential. This formalism was initially introduced in the study of
 non-supersymmetric domain walls in
various dimensions \cite{de,sken}.

For bounce-like solutions, the field $\phi$ tends to different values when $y\rightarrow\pm\infty$. Such solutions can be attained by a double-well potential. In this way, guided by refs. \cite{new,defects-inside,deformed,adalto}, we chosen the superpotential,
\begin{equation}\label{sup}
W_p(\phi)=\frac{p}{2p-1}\phi^{\frac{2p-1}{p}}-\frac{p}{2p+1}\phi^{\frac{2p+1}{p}},
\end{equation}
where the parameter $p$ is an odd integer. The chosen form for $W_p$ was constructed after deforming the $\lambda\phi^4$ model. This choice allows us to obtain well-defined models when $p=1,3,5,...$, where for $p=1$ we get the standard $\phi^4$ potential. For $p=3,5,7,...,$ the potential $V_p$ presents a minimum at $\phi=0$ and two more minima at $\pm 1$. Eq. (\ref{1order}) can be easily solved giving the so called two-kink solutions
\begin{equation}\label{twokink}
\phi_p(y)=\tanh^p\biggl(\frac{y}{p}\biggr).
\end{equation}
From Eq. (\ref{WA}), one find explicitly the solution for $A_p(y)$ as \cite{adalto}
\begin{eqnarray}\label{a}
A_p(y)=-\frac{1}{3}\frac{p}{2p+1}\tanh^{2p}\left(\frac{y}{p}\right)- \frac{2}{3}\left(\frac{p^2}{2p-1}-\frac{p^2}{2p+1}\right)\\\nonumber
\biggl{\{}\ln\biggl[\cosh\left(\frac{y}{p}\right)\biggr]- \sum_{n=1}^{p-1}\frac1{2n}\tanh^{2n}\left(\frac{y}{p}\right)\biggr{\}}
\end{eqnarray}
\begin{figure}[ht!]
\includegraphics[width=13.5cm,height=4.5cm]{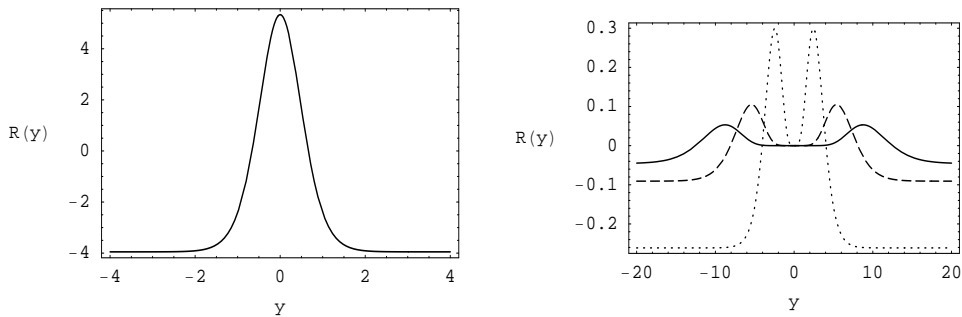}
\caption{\label{curvp}Plots of the solution of the curvature invariant $R(y)$ for $p=1$ on the left. On the right for
$p=3$ (doted line), $p=5$ (dashed line) and $p=7$ (solid line).}
\end{figure}

Note that the exponential warp factor constructed with this function is localized around the membrane and for large $y$ it approximates the Randall-Sundrum solution \cite{rs2}. The spacetime now has no singularities as we get a smooth warp factor (because of this, the model is more realistic) \cite{kehagias}. This can be seen by
calculating the curvature invariants for this geometry. For instance, the Ricci scalar is
\begin{equation}
R=-[8A_p''+20(A_p')^{2}],
\end{equation}
Figs. \ref{curvp}a and \ref{curvp}b shows the Ricci scalar for $p=1,3,5,7$. Note that the Ricci scalar is finite at all points in the bulk, and singularities are absent. Far from the brane $R$ tends to a negative constant, characterizing the $AdS_5$ limit from the bulk. Note that the higher is the parameter $p$, the lower is the constant, interpreted as the inverse of the $AdS$ scale. Fig. \ref{curvp}a is for $p=1$ and shows that the scalar curvature for a non-deformed model presents a maximum at the brane center $y=0$, whereas for higher values of $p$ this maximum is splitted into two smaller peaks that decrease with the increasing of $p$. The presence of regions with positive Ricci scalar can in principle be connected to the capability to trap massive states near to the brane, as we will investigate in the following sections. Also note that for even larger values of $p$ we see that the tendency is the Ricci scalar to approach to zero near to the brane center. In this way we expect the presence of gravity resonances to be more pronounced for lower values of $p$.

\section{Gravity localization and Resonances}

The issue of gravity localization in this class of models was considered by one of us in Ref. \cite{adalto}. As already noted in \cite{gremm}, plane wave solutions of Schrodinger-like equations in the transverse-traceless sector of metric perturbations can present solutions as resonant modes. Such structures were obtained in \cite{csaba,csaba2} when studying gravity localization. Here we review the treatment of metric fluctuations and investigate numerically the presence of resonances with the more refined method.
The stability analysis is performed after perturbing
the metric as follows
\begin{equation}
ds^2=e^{2A(y)}(\eta_{\mu\nu}+h_{\mu\nu})dx^\mu dx^\nu-dy^2.
\end{equation}
Here $h_{\mu\nu}=h_{\mu\nu}(x,y)$ are small perturbations. In the transverse traceless gauge the perturbations turn to ${\bar h}_{\mu\nu}$, and the metric and scalar field fluctuations decouple, resulting in the equation
\begin{equation}
\label{h}
{\bar h}_{\mu\nu}^{\prime\prime}+4\,A^{\prime}
\,{\bar h}_{\mu\nu}^{\prime}=e^{-2A}\,\Box\,{\bar h}_{\mu\nu}.
\end{equation}
 Here $\Box$ stands for the 4-dimensional D'Alembertian. The extra dimension $y$ is turned into a new coordinate $z$, defined by
 \begin{equation}\label{trans1}
 \frac{dz}{dy}=e^{-A_p},\end{equation}
  which makes the metric conformally flat. Also, with
\begin{equation}
{\bar h}_{\mu\nu}(x,z)=e^{ik\cdot x}e^{-\frac{3}{2}A(z)}H_{\mu\nu}(z),
\end{equation}
the equation (\ref{h}) for the metric fluctuations assume the form of a   Schr\"odinger-like equation
\begin{equation}
\label{se}
-\frac{d^2H_{\mu\nu}}{dz^2}+U_p(z)\,H_{\mu\nu}=k^2\,H_{\mu\nu}
\end{equation}
where the potential is given by
\begin{equation}
U_p(z)=\frac32\,A_p^{\prime\prime}(z)+\frac94\,A_p^{\prime2}(z).
\end{equation}

\begin{figure}
\includegraphics[angle=270,width=6.5cm]{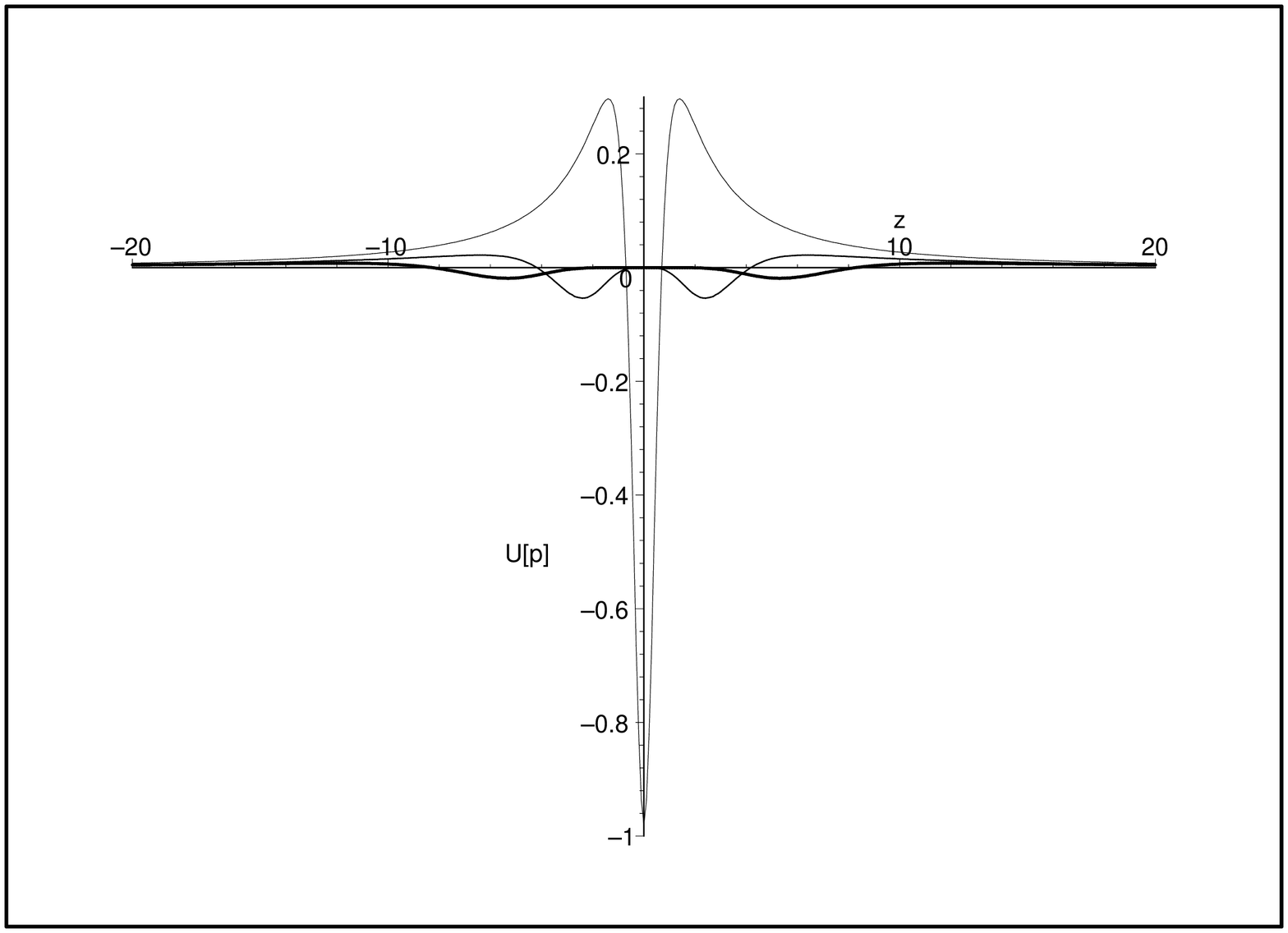}
\includegraphics[{angle=270,width=6.5cm}]{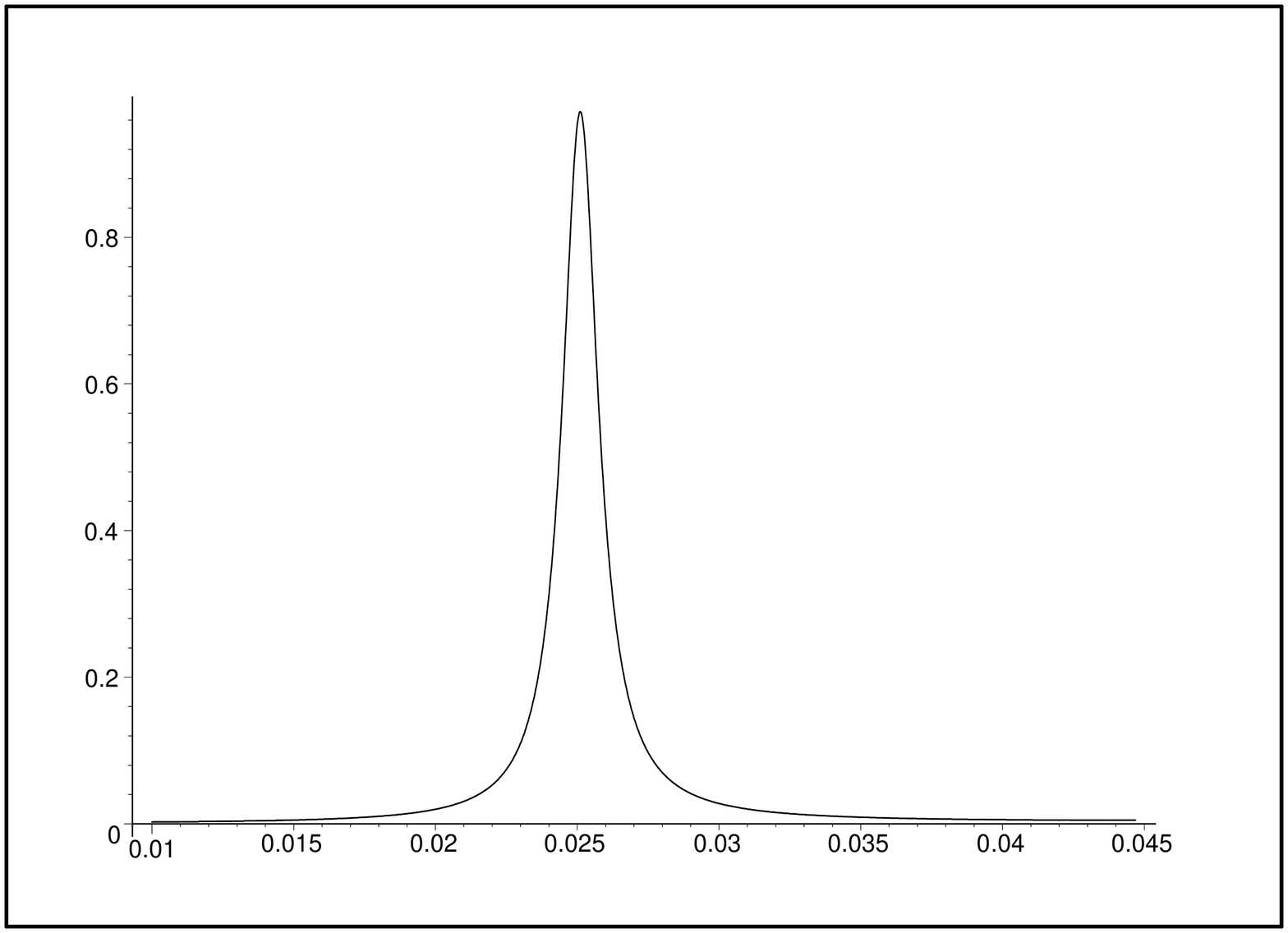}
\includegraphics[{angle=270,width=6.5cm}]{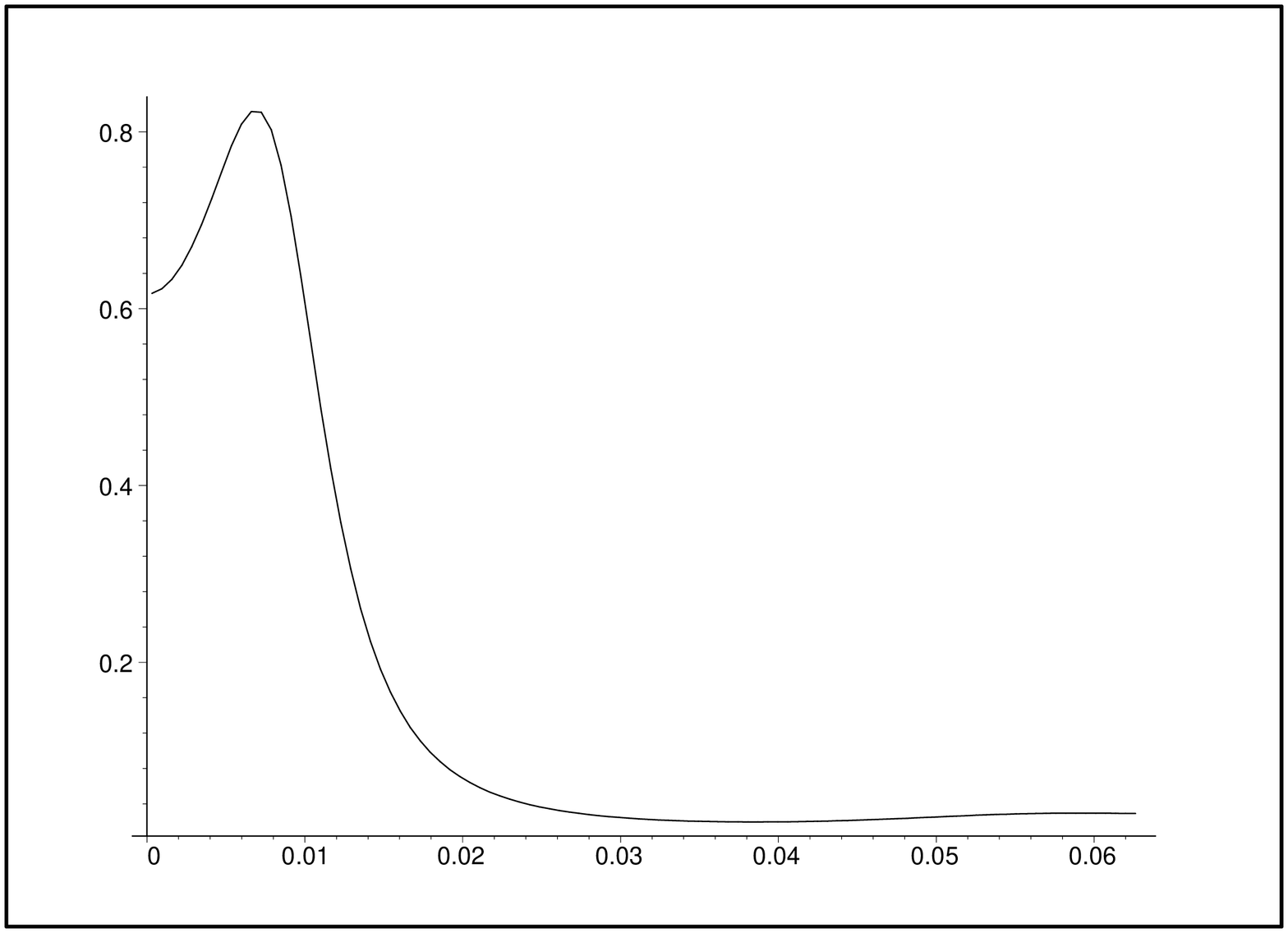}
\includegraphics[{angle=270,width=6.5cm}]{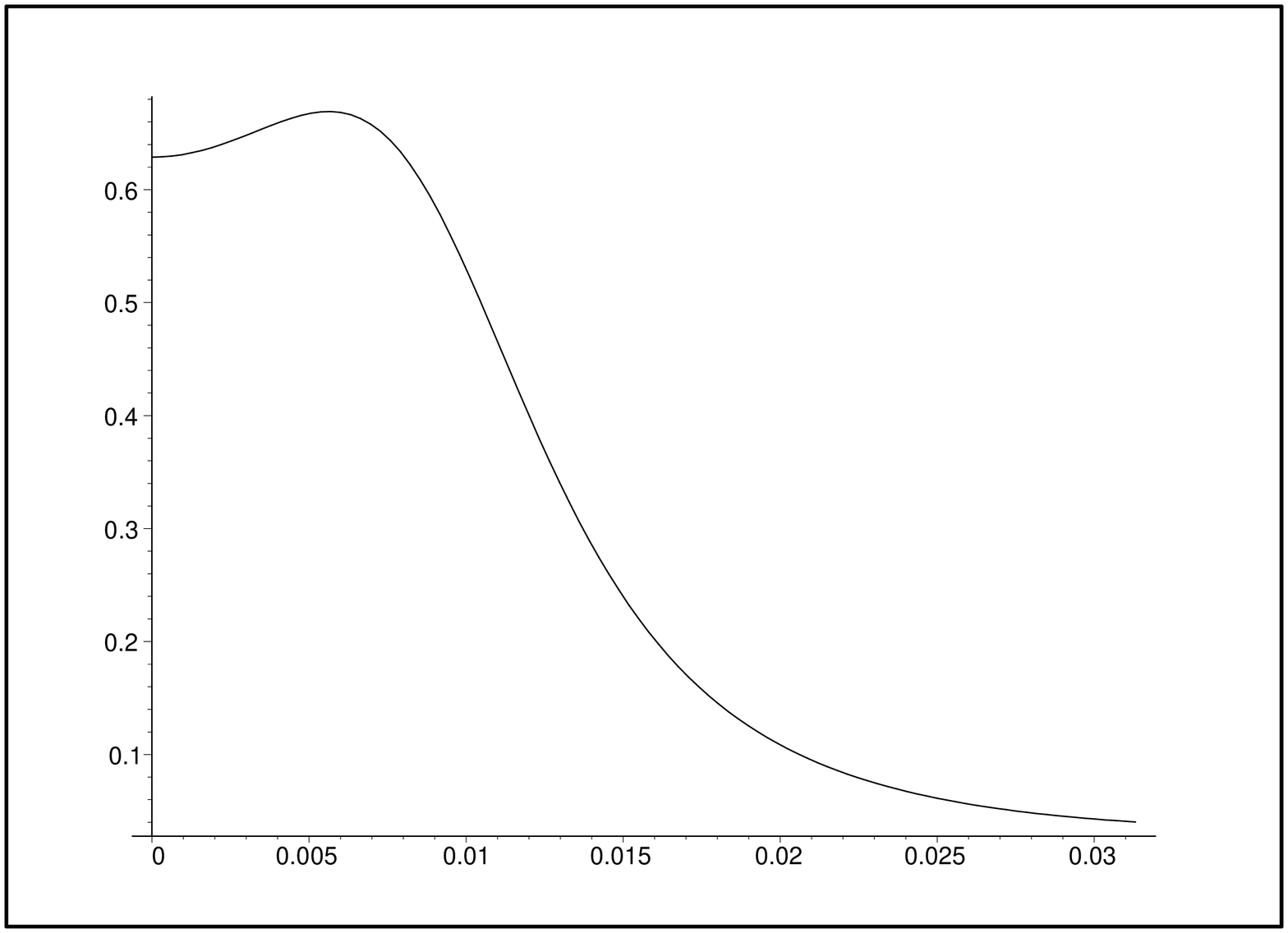}
\caption{(a) (upper left) Plots of the potentials $U_1(z), U_3(z),\, {\rm and}\, U_5(z)$.  Normalized $|H_{\mu\nu}(0)|^2$, as a function of $m$, for $m>0$, showing the resonance peaks for (b) $p=1$ (upper right), $p=3$ (lower left) and $p=5$ (lower right).}
\label{pot-grav}
\end{figure}

Is was already shown \cite{adalto} that the Hamiltonian is positive definite and tachyonic modes are absent and that it is possible to attain an explicit expression for the non-normalized zero-modes, responsible for gravity localization.
Fig. \ref{pot-grav}a shows that the Schrodinger potentials for gravity fluctuations have the form of volcano potentials. This inspired us to investigate the possibility of resonances with the Numerov method \cite{bgl}, identifying resonances as a peak in the normalized squared wavefunctions $|H_{\mu\nu}(0)|^2$ at the brane center.  We considered normalization in a box with ends at $\pm z_{max}$, far enough for the inverse square law $U_p(z)\sim \alpha_p(\alpha_p+1)/z^2$ to be achieved. As gravity localization can be determined by the far region of the potential \cite{csaba,csaba2}, the plot of $z^2U_p(z)$ for $-200<z<200$ gives $\alpha_1=1.483$. This characterizes gravity localization where $U_{grav}(r)$, the gravitational potential  between two unit masses distant $r$ one from the other, reproduces the Newtonian limit for large distances. However, for short distances there is a $1/r^{2\alpha_p}$ correction due to the small massive modes. For $p=1$ this gives $1/r^{2.966}$, close to the Randall-Sundrum $1/r^3$ correction for the Newtonian potential. For larger values of $p$, one needs larger values of $z_{max}$ to achieve the $1/z^2$ region for $U_p$.

For studying resonances, we do not need to pursue to large values of $z_{max}$, since we are interested in the effect where the wavefunction behavior changes abruptly for a particular mass. For our purposes we considered sufficient to consider a normalization procedure with $z_{max}=100$. We found a clear peak for $p=1$, as showed in Fig. \ref{pot-grav}b. We noted that with the increasing of $p$ (Fig. \ref{pot-grav}c and \ref{pot-grav}d), the resonance peaks become broader, showing that branes with smaller values of $p$ are more effective in trapping graviton KK modes. This can also be noted with zero modes, where the modes decay as $1/\sqrt z$ for $z>z_p$ and $z_p$ grow with $p$ (see Ref. \cite{adalto}). For $p=5$ the peak thickness $\Delta m$ is too large for characterizing a resonance. This was expected since the maxima of the Schrodinger potential, firstly pronounced for $p=1$ is reduced considerably for larger values of $p$ (see Fig. \ref{pot-grav}a).

\section{Fermionic zero-mode}

Now we consider the action for a fermion coupled with gravity, in the background given by the brane solution $\phi_p(y)$ from the
previous section. The action is
\begin{equation}
\label{Sferm}
S=\int dx^5\sqrt{g}[\bar{\Psi}\Gamma^{M} D_{M} \Psi - f \bar\Psi\phi_p\Psi],
\end{equation}
where $f$ is the 5-dimensional Yukawa coupling.

Following the literature, firstly we change the variable $y$ to $z$ according to Eq. (\ref{trans1}).
In the new set of variables the metric is conformally plane, and the gamma matrices can be rewritten as
$\Gamma^{\mu}=e^{-A_p}\gamma^{\mu}\;,\;\Gamma^5=e^{-A_p}\gamma^5$. The covariant derivatives are
\begin{equation}
D_{\mu}=\partial_{\mu}+\frac{\partial_z A_p}{2}\gamma_{\mu}\gamma^5\;,\;\;\;\;D_5=\partial_5.
\end{equation}
The former expressions allow us to write the equation of motion as
\begin{equation}
[\gamma^\mu\partial_\mu+\gamma^5(\partial_z+2\partial_zA_p)+f\phi_p e^{A_p}]\Psi(x,z)=0.
\end{equation}
Now we perform a chiral decomposition of the 5-dimensional spinor $\Psi$ as
\begin{equation}
\Psi(x,z)=\sum_n[\psi_{Ln}(x)\alpha_{Ln}+\psi_{Rn}(x)\alpha_{Rn}(z)].
\end{equation}
The massive modes from the spinor $\Psi$ living on the brane must connect both chiralities, satisfying the equations
\begin{equation}
\gamma^{\mu}\partial_{\mu}\psi_{Ln}(x)=m\psi_{Rn}(x)\,,\,\,\gamma^{\mu}\partial_{\mu}\psi_{Rn}(x)=m\psi_{Ln}(x).
\end{equation}
From the relations  $\gamma^5\psi_{Ln}(x)=-\psi_{Ln}(x)$ and $\gamma^5\psi_{Rn}(x)=\psi_{Rn}(x)$, we find two coupled equations for  $\alpha_{Ln}(z)$ and $\alpha_{Rn}(z)$:
\begin{equation}
\label{eom_psi}
[\partial_z+2\partial_zA_p + f\phi_pe^{A_p}]\alpha_{Ln} (z)= m\alpha_{Rn}(z),
\end{equation}
\begin{equation}
\label{eom_psi2}
[\partial_z+2\partial_zA_p - f\phi_pe^{A_p}]\alpha_{Rn} (z)=- m\alpha_{Ln}(z).
\end{equation}

Now we investigate the possibility of localized fermionic zero massive modes. For $m=0$, Eq. (\ref{eom_psi}) for $\alpha_{Ln}(z)$, reduces to
\begin{equation}
2A'_p \alpha_{Ln}(z)+\alpha_{Ln}'(z)+fe^{A_p}\phi_p\alpha_{Ln}(z)=0.
\end{equation}
If we turn back to the variable $y$ we can write
\begin{equation}
2A'_p \alpha_{Ln}(y)+\alpha_{Ln}'(y)+f\phi_p\alpha_{Ln}(y)=0.
\end{equation}
with solution
\begin{equation}\label{sol}
\alpha_{Ln}(y)=e^{-\int_0^y dy' [f\phi_p+2 A_p(y')]}.
\end{equation}
In this solution we clearly note the contribution of the internal structure from the membrane. Remembering that
$\phi_p$ and $A_p$ depend on odd integer numbers, we will see that the value of $p$ will be determinant in order to obtain a finite solution.
With the explicit expressions for $A(y)$ and $\phi_p(y)$ in Eq. (\ref{sol}), and following Ref. \cite{kehagias}, we where able to find the relation
\begin{equation}\label{rel}
f>\frac{8p}{-3+12p^2}.
\end{equation}
between the coupling constant $f$ and the parameter $p$ for $\alpha_{Ln}(y)$ to be finite.

\begin{figure}
\includegraphics[width=7.6cm,height=5.7cm]{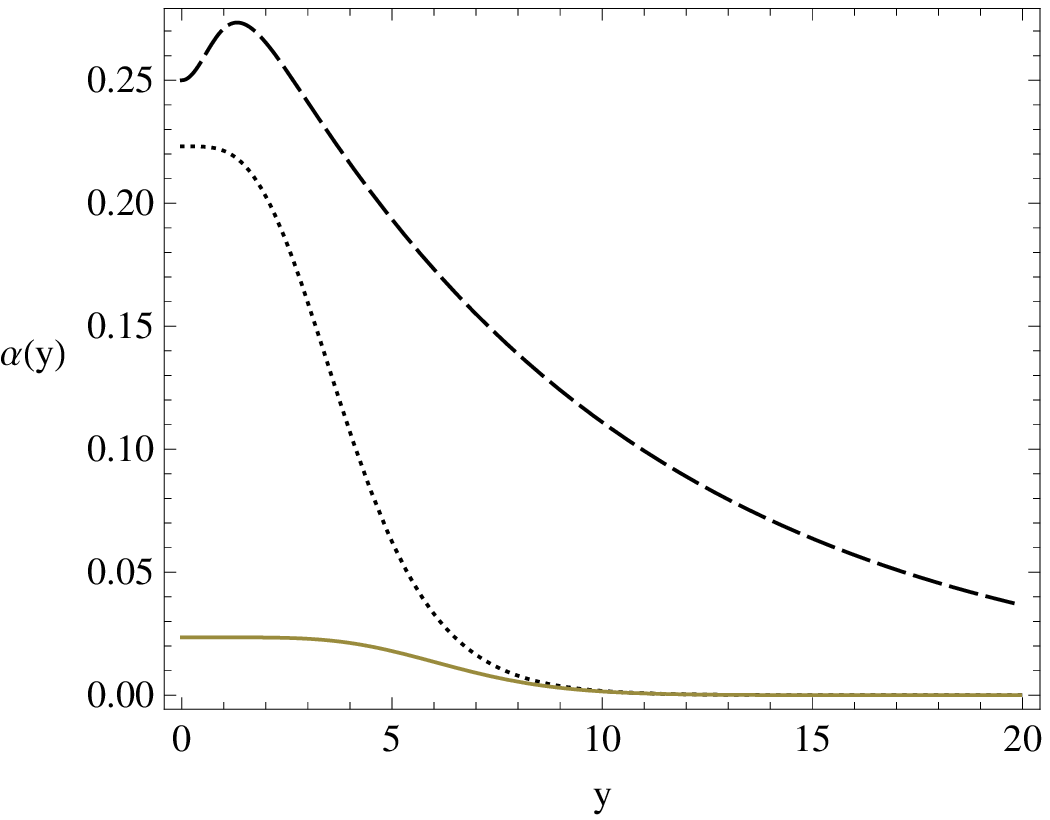}
\includegraphics[width=7.3cm,height=5cm]{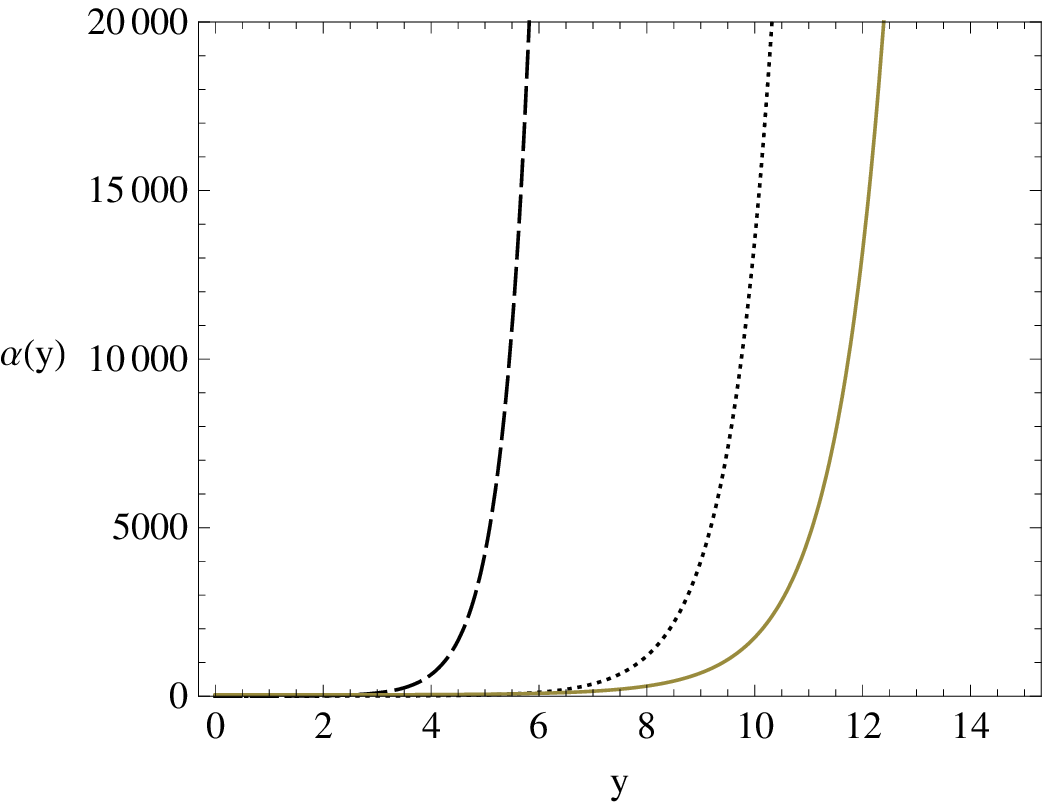}
\caption{(a) Plots of zero-mode $\alpha_{Ln}$ (left) and $\alpha_{Rn}$ (right) with $f=1$ for $p=1$ (dashed line), $p=3$ (doted line) and $p=5$ (solid line).}
\label{ne} 
\end{figure}

As an example, we consider the particular case where $p=3$ in order to determinate the effects of the deformations on the issue of the localization of the solution. In this case, Eq. (\ref{sol}) can be written as
\begin{equation}
\alpha_{Ln}(y)=e^{-\frac{3}{2}f sech(\frac{y}{3})^2+\frac{2}{35}tanh(\frac{y}{3})^2[-6-3 tanh(\frac{y}{3})^2+5
tanh(\frac{y}{3})^4]}cosh\left(\frac{y}{3}\right)^{\frac{24}{35}-3f}.
\end{equation}
The exponential part of the solution above tends to a constant value in regions far from $y=0$ for any value of $f$. Therefore, the exponent $\frac{24}{35}-3f$ determines the form of the solution when $y\rightarrow\pm\infty$. In order to obtain a finite solution a following condition must be obeyed
\begin{equation}
\frac{24}{35}-3f<0.
\end{equation}
Analyzing the solution for $A_p(y)$ in (\ref{a}), we note that the factor $24/35$ in the solution for $p=3$ results from the term,
\begin{equation}
\frac{4}{3}\left(\frac{p^2}{2p-1}-\frac{p^2}{2p+1}\right).
\end{equation}
Simplifying the term above, we obtain the general relationship from Eq. (\ref{rel}) between $f$ e $p$ that must be satisfied for all solutions for $\alpha_{Ln}(y)$ to be finite. It is important to point out that the relation from Eq. (\ref{rel}) between $f$ e $p$ results from the coupling of the fermion $\Psi$ with the particular kind of membrane solution, introduced in the action as $f\overline{\Psi}\phi_p\Psi$.
Fig. (\ref{ne}a) shows the solution $\alpha_{Ln}(y)$ for $p=1,3,5$ for $y>0$. Note from the plots that the solutions asymptote to zero for large $y$, as required for localized solutions. Also note that small values of $p$ shows higher peaks, characterizing better localization.

For $\alpha_{Rn}(y)$, Eq. (\ref{eom_psi2}) for $m=0$, after turning to the variable $y$, reduces to
\begin{equation}
2A'_p \alpha_{Rn}(y)+\alpha_{Rn}'(y)-f\phi_p\alpha_{Rn}(y)=0,
\end{equation}
with solution
\begin{equation}
\label{sol2}
\alpha_{Rn}(y)=e^{\int dy' [f\phi_p(y')-2 A_p(y')]}.
\end{equation}
From the solutions $A(y),\phi_p(y)$ we conclude that there is no localized right chiral zero-mode for $f$ satisfying Eq. (\ref{rel}).
This can be seen explicitly in Fig. (\ref{ne}b) where we plot solutions from Eq. (\ref{sol2}) for $p=1,3,5$.  Note from the figure that the $p=1$ case diverges for lower values of $y$ in comparison with larger values of $p$. At least for zero modes we conclude that lower values of $p$ favor localization of left chiral modes and disfavor localization of right chiral modes. We expect this to be also valid for massive modes. Further we will confront our findings for the zero-modes with the general calculations for the massive modes.

Now to complete our analysis for chiral zero-modes we must verify the normalizability of our solutions. Decomposing the
$y$-dependent part from the action for free fermions given by the first term from Eq. (\ref{Sferm}) we get
\begin{eqnarray}
S_{free}&=&\int d^4x\int_{-\infty}^{+\infty}dy\sqrt{g}\bar{\Psi}(x,y)\Gamma^A D_A \Psi(x,y)\\\nonumber &=&\int_{-\infty}^{+\infty}dy e^{3A_p(y)}|\alpha(y)|^2\int
d^4x \bar{\psi}(x)\gamma^{\mu}\partial_{\mu}\psi(x).
\end{eqnarray}
This expression shows that only left zero-mode chiral solutions are normalizable, as lead to a finite integral in $y$ \cite{kehagias}.

\section{Fermionic massive modes}
Now to complete our investigation of the presence of spinorial fields in the 4-dimensional membrane, we consider the Dirac massive equation. Since we are interested in massive localized states, we transform the equation of motion for fermions in a Shr\"odinger-like equation, a well-known procedure also used in our study of gravity localization in Sec. III. For fermions in a domain wall there is a similar analysis \cite{peter}. Double walls were studied in  \cite{alejandra}.

With the transformations
$\alpha_{Ln}(z)=\overline{\alpha}_{Ln}(z) e^{-2A_p}$ and $\alpha_{Rn}(z)=\overline{\alpha}_{Rn}(z) e^{-2A_p}$, Eqs. (\ref{eom_psi}) and (\ref{eom_psi2}) result in
\begin{equation}\label{schro}
[-\partial^2_z +V_p^L]\overline{\alpha}_{Ln}(z)=m^2\overline{\alpha}_{Ln}(z),
\end{equation}
\begin{equation}\label{schro2}
[-\partial^2_z +V_p^R]\overline{\alpha}_{Rn}(z)=m^2\overline{\alpha}_{Rn}(z),
\end{equation}
where the Schr\"odinger potentials are
\begin{eqnarray}
V_p^L&=&- f\partial_z\phi_p e^{A_p}- f\phi_p e^{A_p}\partial_z A_p+f^2\phi_p^2 e^{2A_p}\\
V_p^R&=& +f\partial_z\phi_p e^{A_p}+ f\phi_p e^{A_p}\partial_z A_p+f^2\phi_p^2 e^{2A_p}.
\end{eqnarray}
Due to the change on variables from $y$ to $z$ we have no explicit form for the potentials $V_p^L$ and $V_p^R$. From the numerically known potential we can use the Numerov numeric method \cite{bgl}
to solve the Schr\"odinger equations.

\begin{figure}
\includegraphics[width=7.0cm,height=4.3cm]{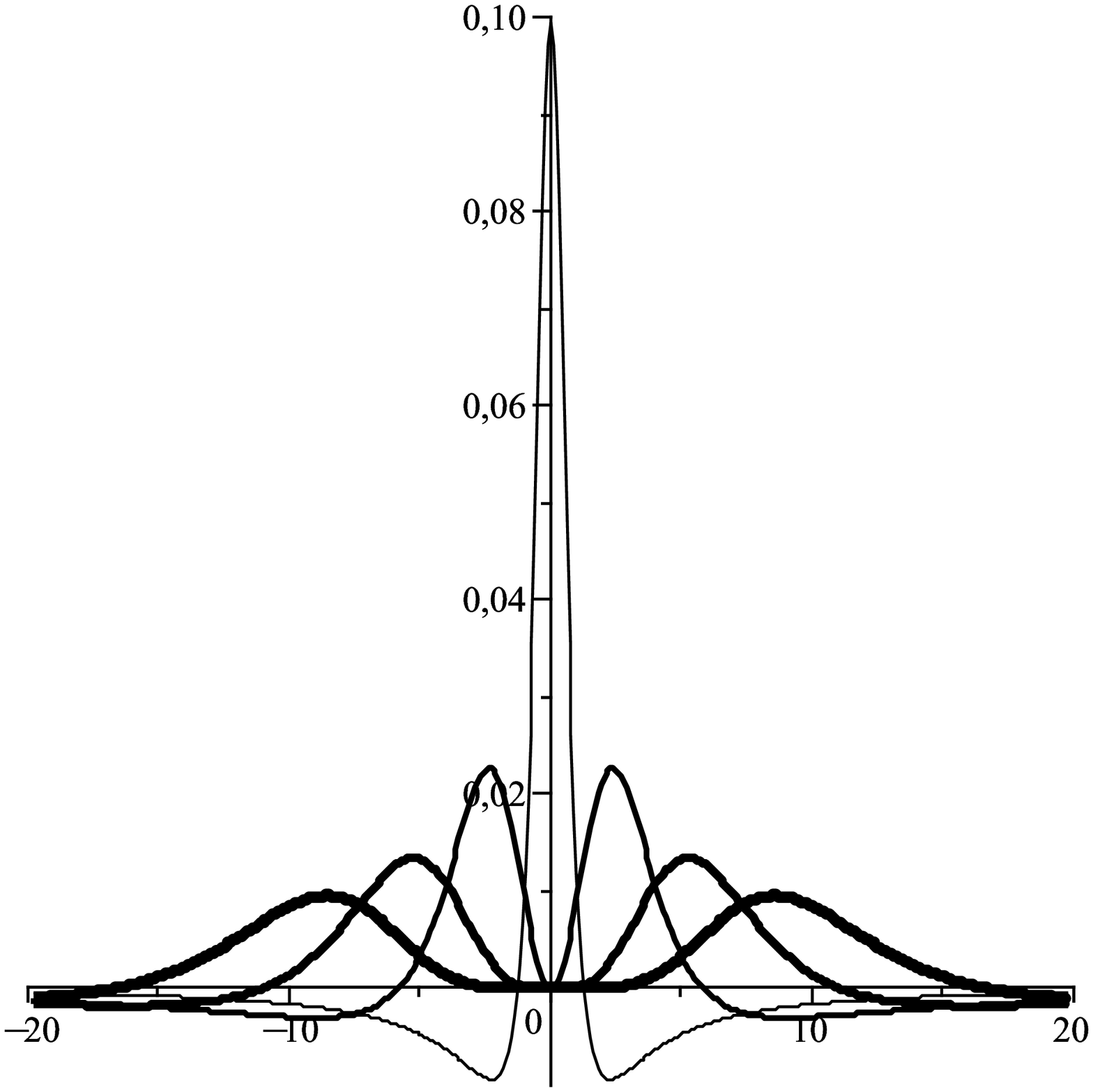}
\includegraphics[width=7.0cm,height=4.3cm]{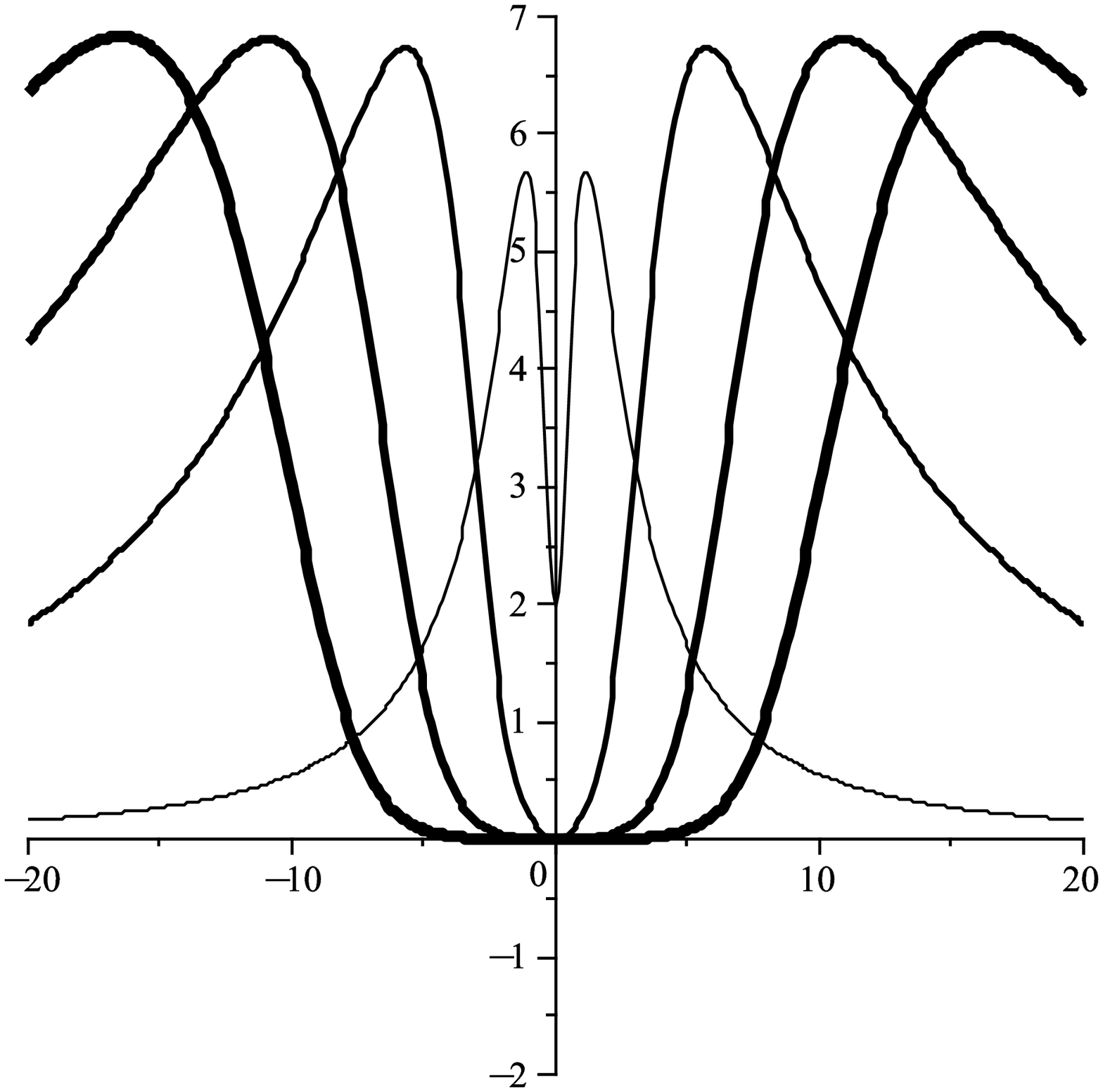}
\caption{\label{neg}Plots of $V_p^R(z)$ (right chiral mode potential) with (a) $f=0.1$ (left) and (b) $f=2$ (right) fixed. Curves correspond to $p=1$ (tinner line), $3$, $5$ and $7$ (thicker line).}
\end{figure}

Fig. (\ref{neg}) shows the potentials $V_p^R$ for $p=1,3,5$ and fixed $f=0.1$ and $f=2$. Note from the figure that the potential for $p=1$ is qualitatively different from $p=3,5$. For $p=1$ and $f=0.1$ the potential has a maximum at the center of the brane, and the structure of the potential shows that there is no resonance. For $p=3,5$, there appears a structure of two peaks separated by a local minimum at $z=0$. The case where $f=2$ is more interesting, where there appears a hole for $p=1$. For $p=3,5,7$ the height of the peaks is roughly the same, but the peaks are more distant for larger values of $p$, showing that the internal structure of the brane favor the appearance of resonances.

\begin{figure}
\includegraphics[width=7.0cm,height=4.3cm]{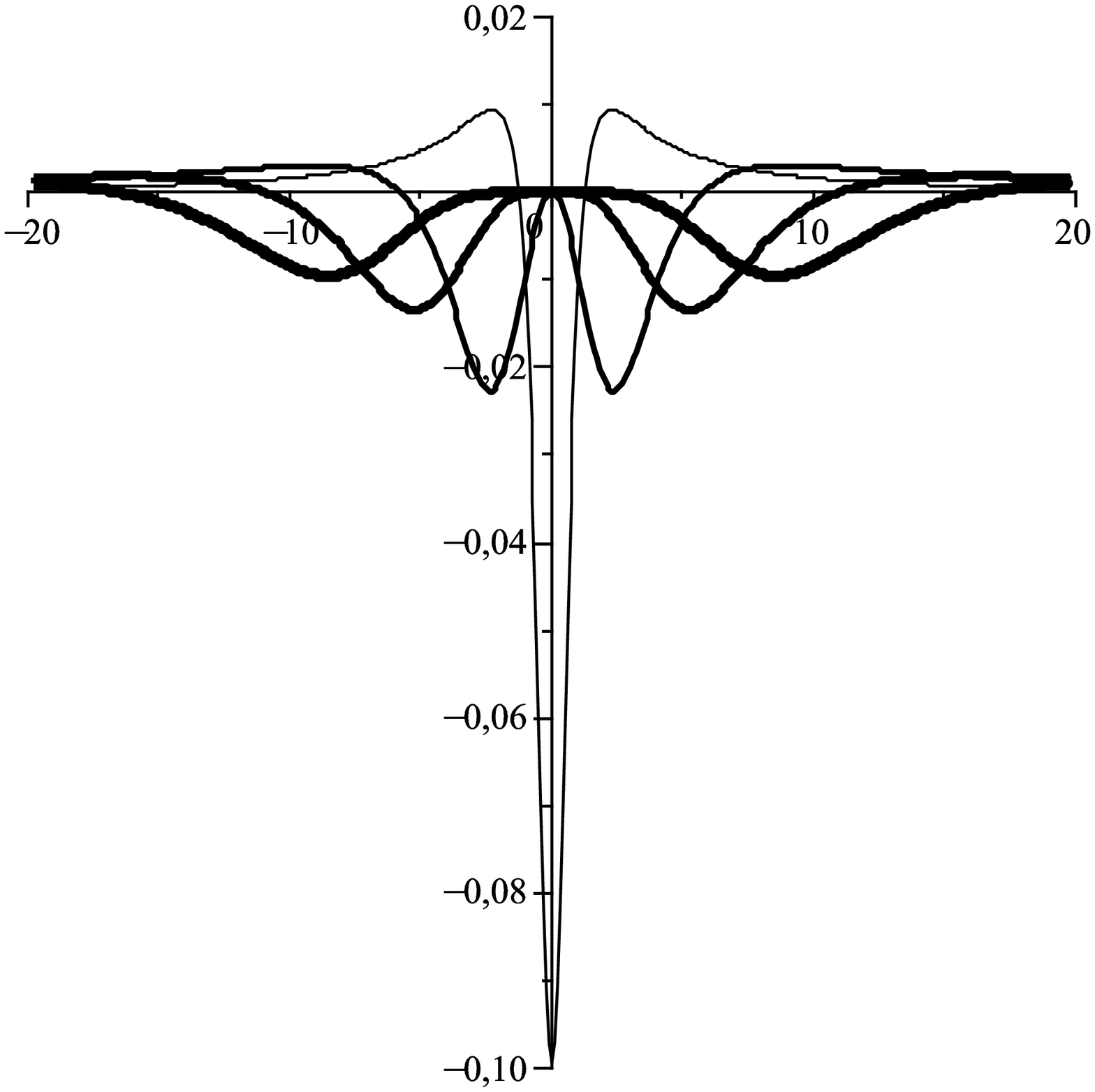}
\includegraphics[width=7.0cm,height=4.3cm]{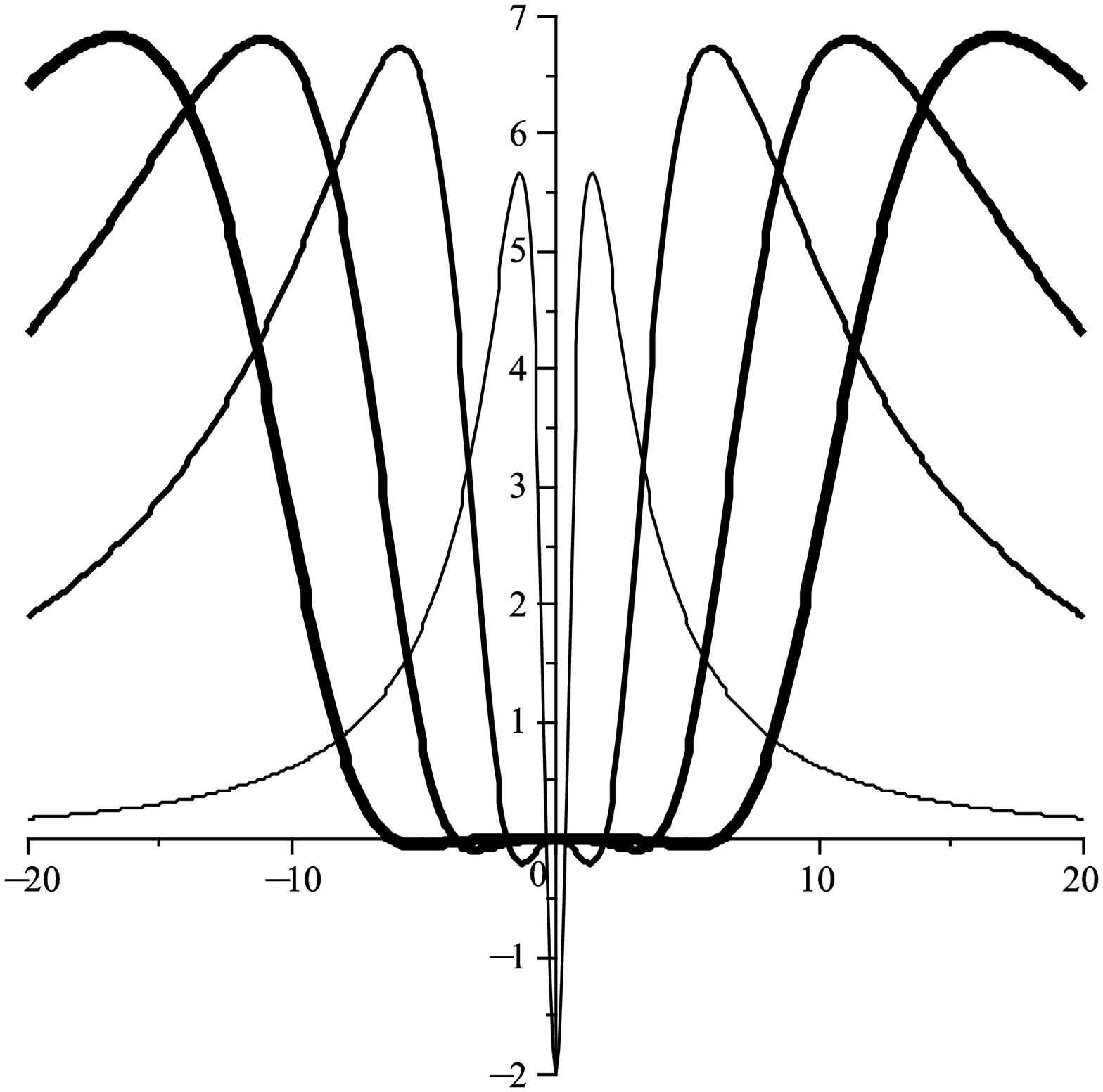}
\caption{\label{pos}Plots of $V_p^L(z)$ (left chiral mode potential) with (a) $f=0.1$ (left) and (b) $f=2$ (right) fixed. Curves correspond to $p=1$ (tinner line), $3$, $5$ and $7$ (thicker line).}
\end{figure}

Fig. (\ref{pos}) shows the potential  $V_p^L$ for several values of $p$. In this case there is only one minima for
$p=1$, where the potential is negative. This shows that a brane with this potential has a bound state, the zero-mode already studied.  For larger values of $p$, this central minimum is separated into two minima located far from the brane for larger values of $p$. We note that these minima are less pronounced for larger values of $p$, disfavoring the localization of the zero-modes. This agrees with Fig. \ref{ne}a, where we see that the solution for the zero-mode $\alpha_{Ln}(y)$ for $p=1$ is characterized by a higher peak in comparison to what found for larger values of $p$. Fig. (\ref{pos}) also shows that the potential is characterized by two positive maxima. This shows that for energies between zero and this maxima one can investigate the presence of resonances. Note also that for $p\ge 3$ and for small values of $f$, say $f=0.1$, with the increasing of $p$, the height of the two peaks from the potential decrease and are more apart one from the other (see Fig. (\ref{pos}a)). This is a similar effect was also noted in the Sec. III of this paper, when studying the Shr\"odinger potential in the transverse-traceless sector of metric fluctuations. In this case we would expect that higher values of $p$ are less effective to trap the resonant massive modes. However, such small values of $f$ do not satisfy the inequality (\ref{rel}), and zero-modes are non-normalizable. On the contrary, for larger values of $f$, for instance $f=2$, the height of the two peaks are almost constant for $p\ge 3$ and are also more apart one from the other (see Fig. (\ref{pos}b)). Such effect where also found in studying gravity fluctuations in ref. \cite{fase}, where such characteristics are compared to the phase transition of complete wetting in condensed matter systems. In our case this lead us to expect that higher values of $p$ are more effective to trap the resonant massive modes, increasing their lifetimes.

\begin{figure}
\includegraphics[angle=-90,width=7cm]{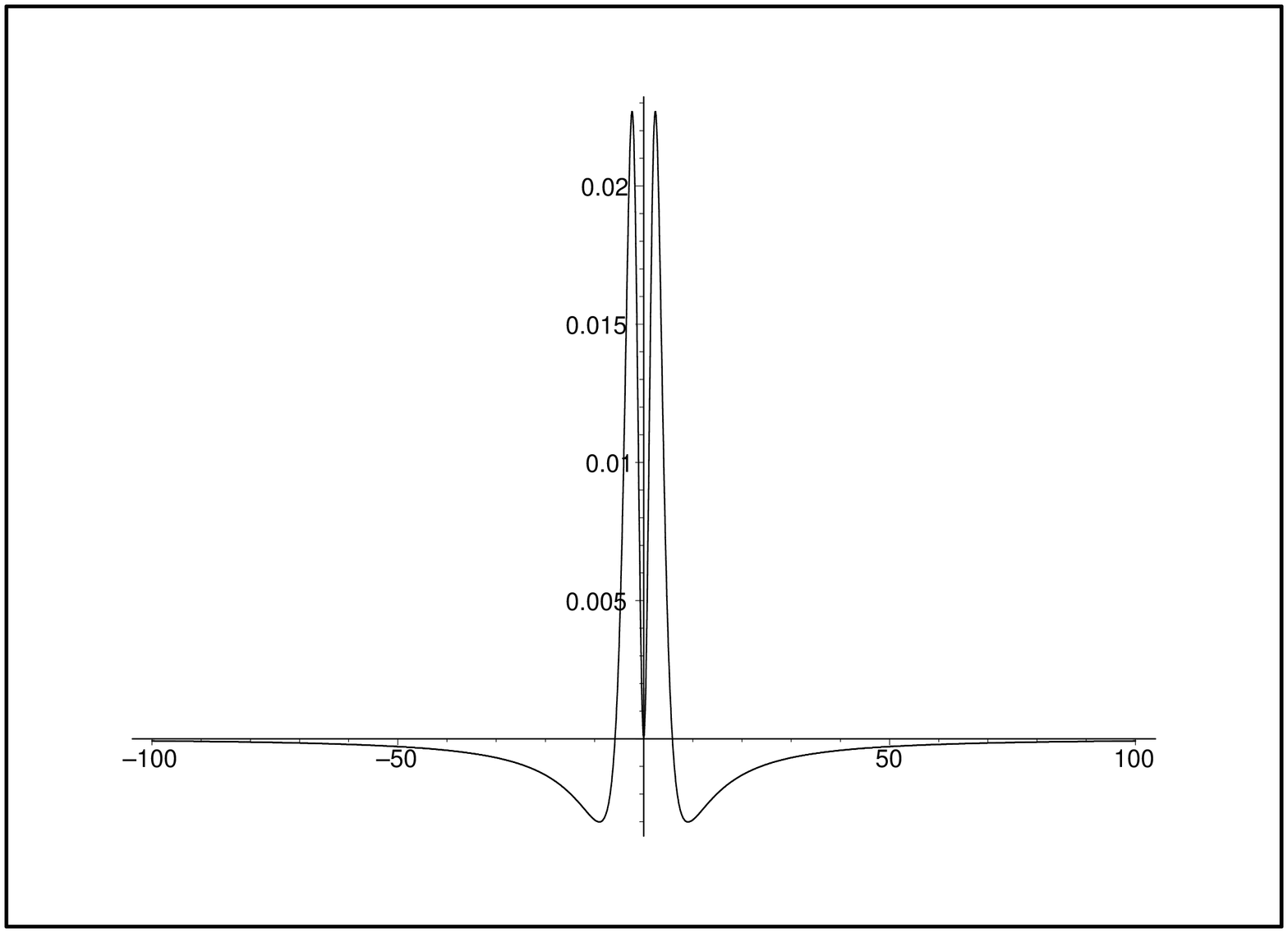}
\includegraphics[angle=-90,width=7cm]{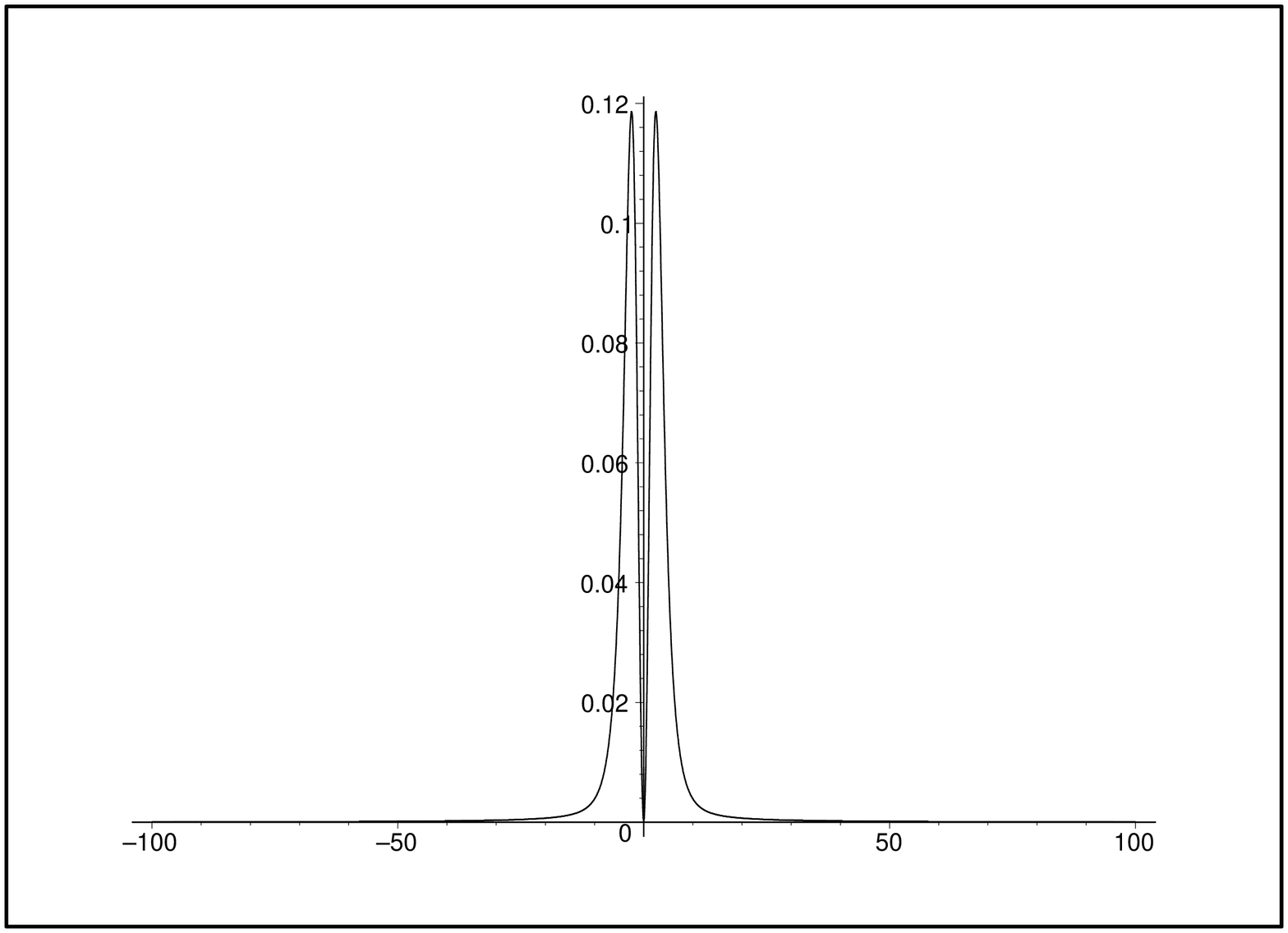}
\includegraphics[angle=-90,width=7cm]{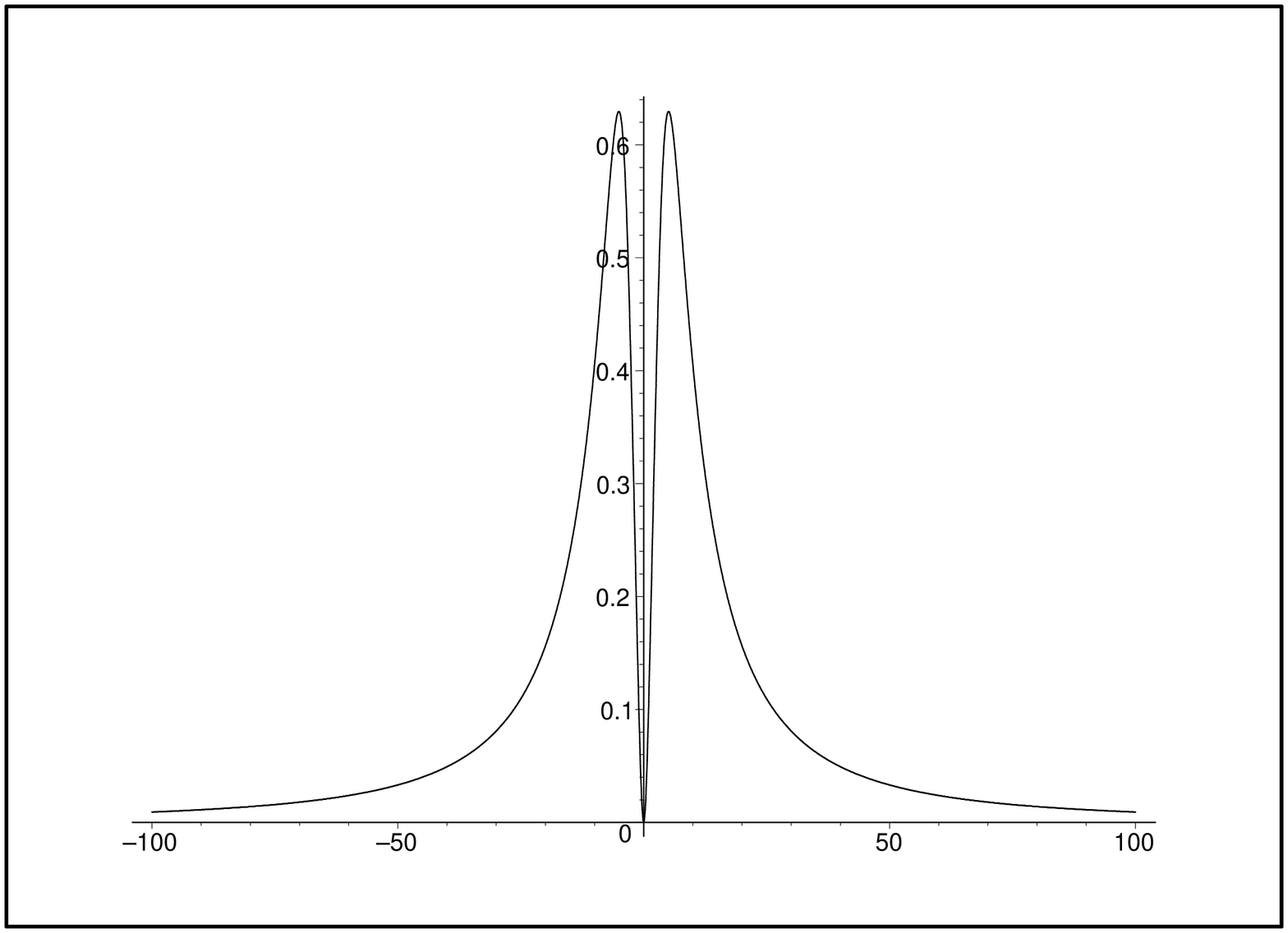}
\includegraphics[angle=-90,width=7cm]{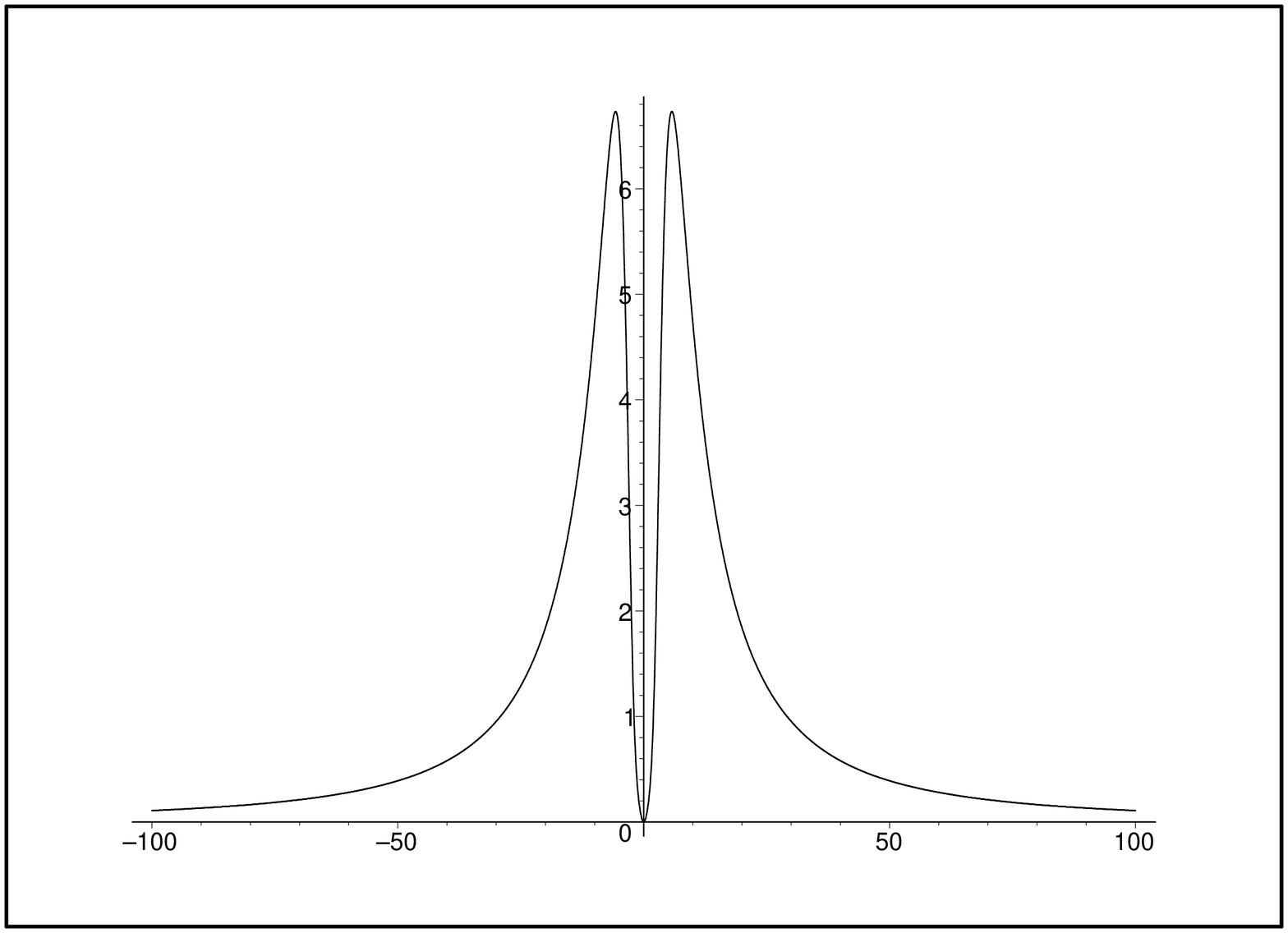}
\caption{\label{VR_f}Plots of $V_p^R(z)$ with $p=3$, for $f=0.1$ (upper left), $0.5$ (upper right) , $1.1$  (lower left), $2$ (lower right).}
\end{figure}

We can analyze more closely the influence of the 5-dimensional Yukawa coupling. As an specific example we chosen to fix $p=3$ and varied the parameter $f$ for right-chiral fermions. Fig. \ref{VR_f} shows that for $f=0.1$, where inequality (\ref{rel}) is violated, the potential is qualitatively different for the righer values of $f$ analyzed, assuming negative values far from the brane.  For the larger values of $f$ presented in the figure,  the position of the two maxima is roughly the same, independent of $f$. The height of the maxima increase faster with $f$, showing a tendency for more resonances to appear as the coupling $f$ increases.

Note that for all potentials in Figs. (\ref{neg}) and (\ref{pos}), the Shr\"odinger potential asymptotes to zero, and there is no gap in the spectra. The potentials for both chiralities show significant changes when the parameter $p$ changes from 1 to larger values. The appearance of internal structure is the important point and can reveal other physical aspects related to fermion localization. In order to investigate the confining of the massive chiral modes, we must solve Eqs. (\ref{schro}) and (\ref{schro2}). It is easy to see that for massive modes that equations can be rewritten as
\begin{eqnarray}\label{fqm}
\emph{Q}\emph{Q}^+\overline{\alpha}_{Rn}=[\partial_z + f\phi_p e^{A_p}][-\partial_z + f\phi_p
e^{A_p}]\overline{\alpha}_{Rn}=m^2\overline{\alpha}_{Rn},\\\nonumber \emph{Q}^+\emph{Q}\overline{\alpha}_{Ln}=[-\partial_z + f\phi_p e^{A_p}][\partial_z +
f\phi_p e^{A_p}]\overline{\alpha}_{Ln}=m^2\overline{\alpha}_{Ln},
\end{eqnarray}
corresponding to a supersymmetric quantum mechanics scenario. Written the equations of motion in this form, tachyonic modes are clearly forbidden.
\begin{figure}
\includegraphics[width=14cm,height=4.7cm]{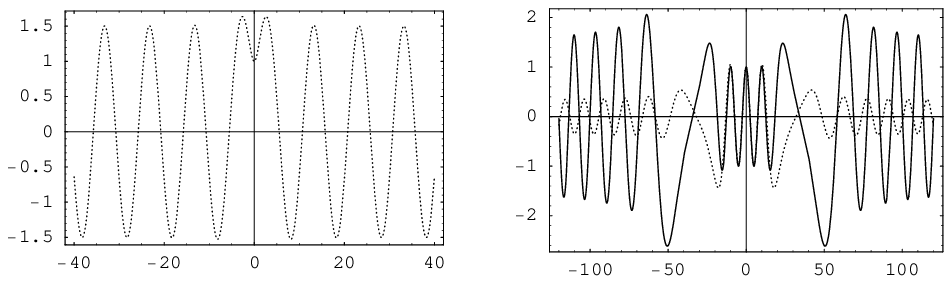}\caption{\label{wave1}Plots of $\overline{\alpha}_{Rn}$  for (a) $p=1$ (left), and (b) $p=9$ (right, doted line), $p=11$ (right, solid line). We fix $f=1$ and $m=0.4$.}
\end{figure}

\begin{figure}
\includegraphics[width=14cm,height=4.7cm]{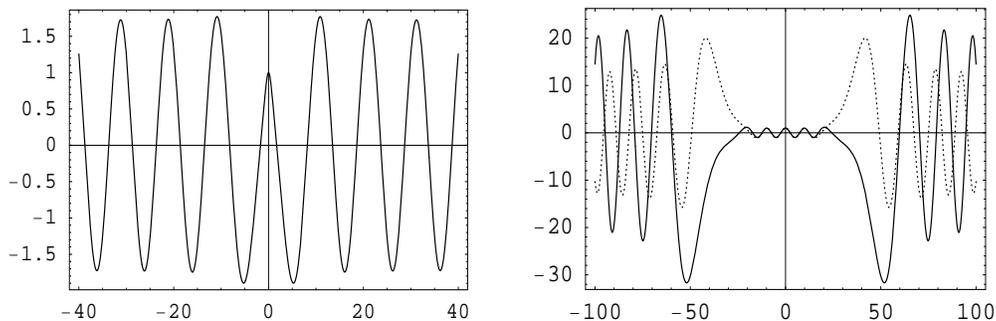}\caption{\label{wave2}Plots of $\overline{\alpha}_{Ln}$  for $p=1$ (left), and (b) $p=9$ (right, doted line), $p=11$ (right, solid line). We fix $f=1$ and $m=0.4$.}
\end{figure}
We numerically solve Eqs. (\ref{schro})-(\ref{schro2}) and plot the solutions for $\overline{\alpha}_{Rn}$ and $\overline{\alpha}_{Ln}$ for some values of $p$ and $f$ in Figs. (\ref{wave1}) and (\ref{wave2}), respectively. A method for solving such equations can be found in Refs. \cite{bgl,ca}. In each figure we show the solutions for $p=1$ on the left representing kink models already considered in the literature \cite{peter,alejandra}, whereas on the right we shown the solutions on the deformed walls. In these figures we considered higher values of $p$ where the difference between the background treated here and the $p=1$ is clearer. Note that the solutions show a transition region for both chiralities near the center of the brane. Far from the brane the solutions have the character of plane waves, signifying that the fermion escapes from the brane.

\section{Fermionic Resonances and Dirac fermions}

The changing of variables that produced the Shr\"odinger equations presented in Eq. (\ref{fqm}), lead us to adopt a quantum mechanical interpretation for $\bar\alpha_{Ln}$ and $\bar\alpha_{Rn}$. One importance of studying resonances is connected to their information of the coupling of massive modes and the brane, illustrating how the mechanism of fermion trapping is being processed. In our case, we can interpret   $|N\overline{\alpha}_\pm(z)|^2$  as the probability for finding the massive mode in the position $z$, with $N$ a normalization constant. In this way, calculating $P(m)\equiv|N\overline{\alpha}_\pm(0)|^2$ as a function of the mass $m$, we are able to detect resonant modes as large peaks in the plot $P(m)$ {\it versus} $m$.

First of all we investigate right chirality, where there is no zero-mode. Fig. \ref{neg}a shows that for the case $p=1$ and small $f$ there is no local minima for the potential, and resonances are absent. The presence of such minima for $p\ge 3$ as depicted in Fig. \ref{neg}b shows that resonances possibly exist for a wider range of $f$. In particular for $f=2$, Fig. \ref{neg}b shows that the maxima of the potentials for $p=3,5,7...$ are nearly the same, as well as the local minima. However, the region in the potential near the local minima is broader for larger values of $p$. This signals that for a large value of $f$, larger values of $p$ are more effective in trapping the fermionic massive modes in comparison to smaller values of $p$. This must be compared to the richer structure of the energy density of the branes for larger values of $p$, as noted in \cite{adalto}.
\begin{figure}
\includegraphics[angle=-90,width=4.5cm]{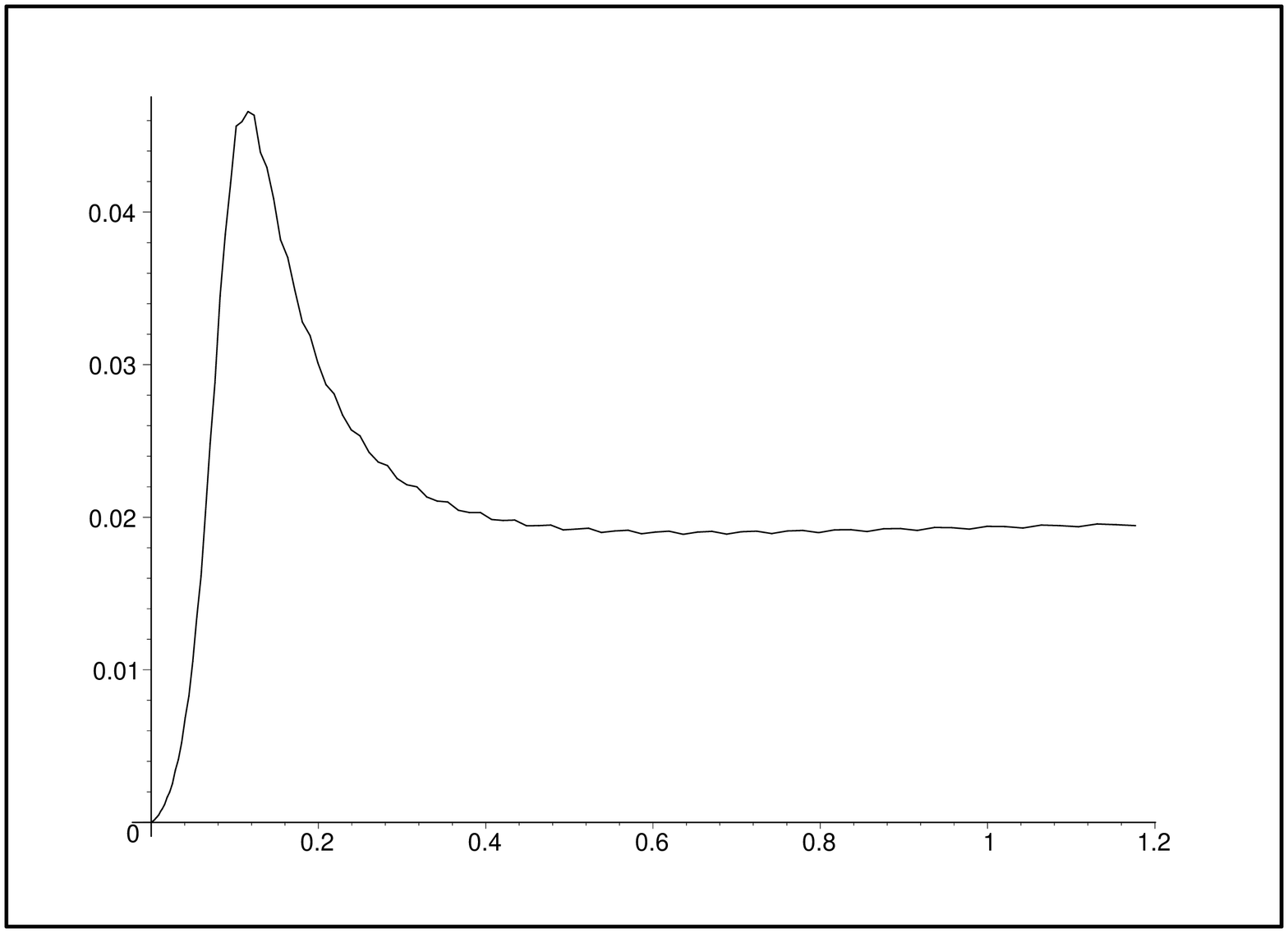}
\includegraphics[angle=-90,width=4.5cm]{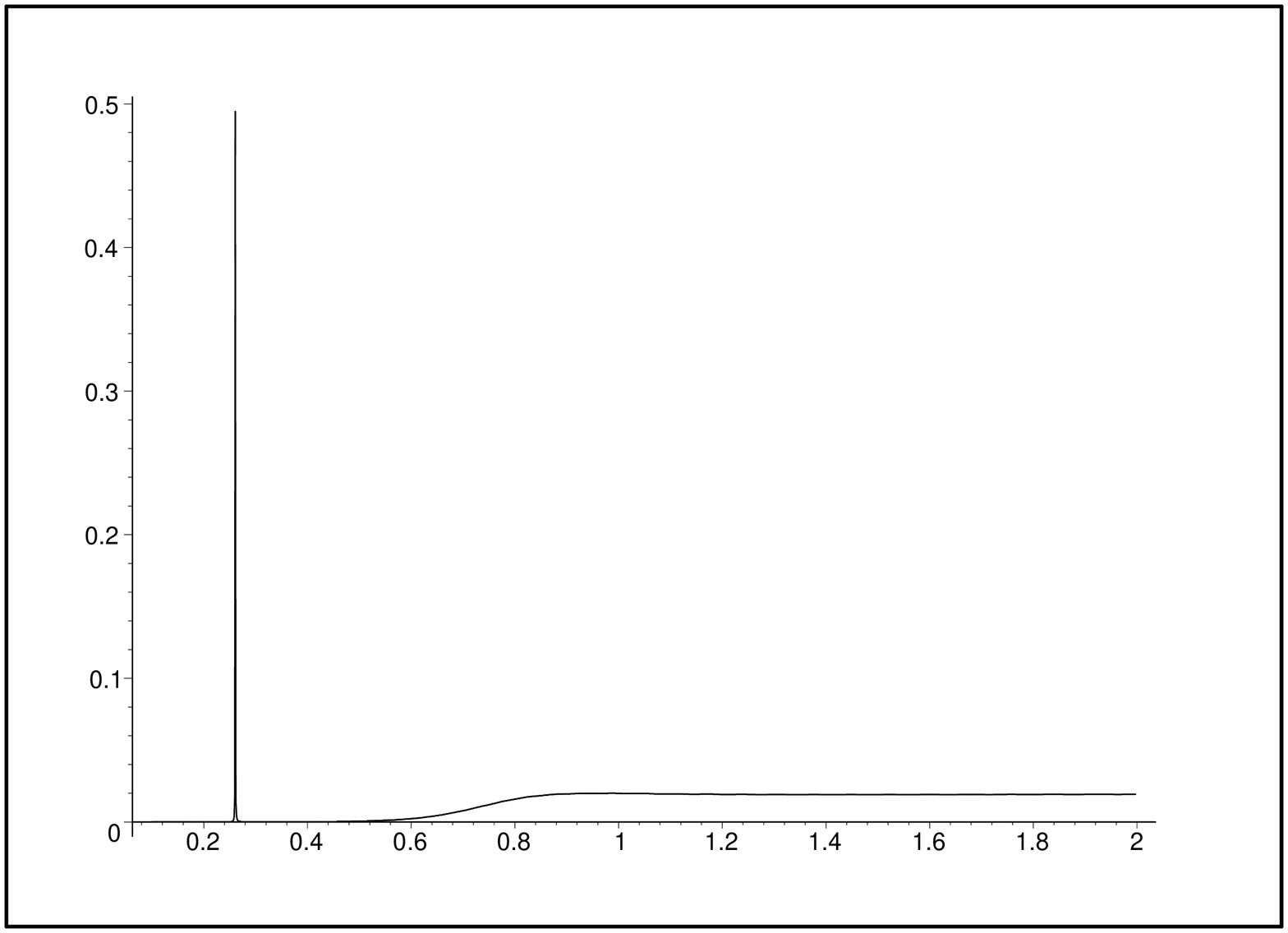}
\includegraphics[angle=-90,width=4.5cm]{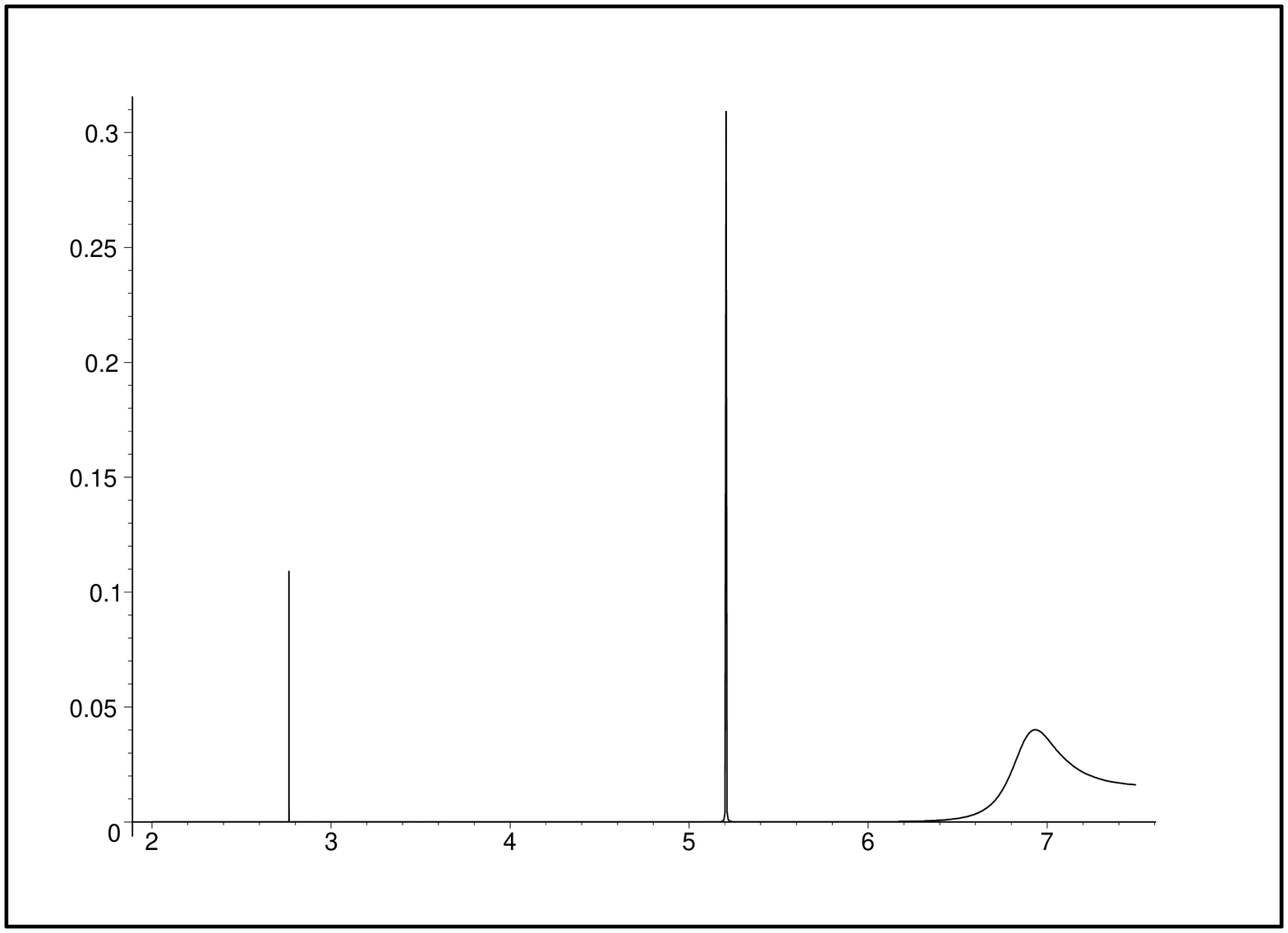}
\includegraphics[angle=-90,width=4.5cm]{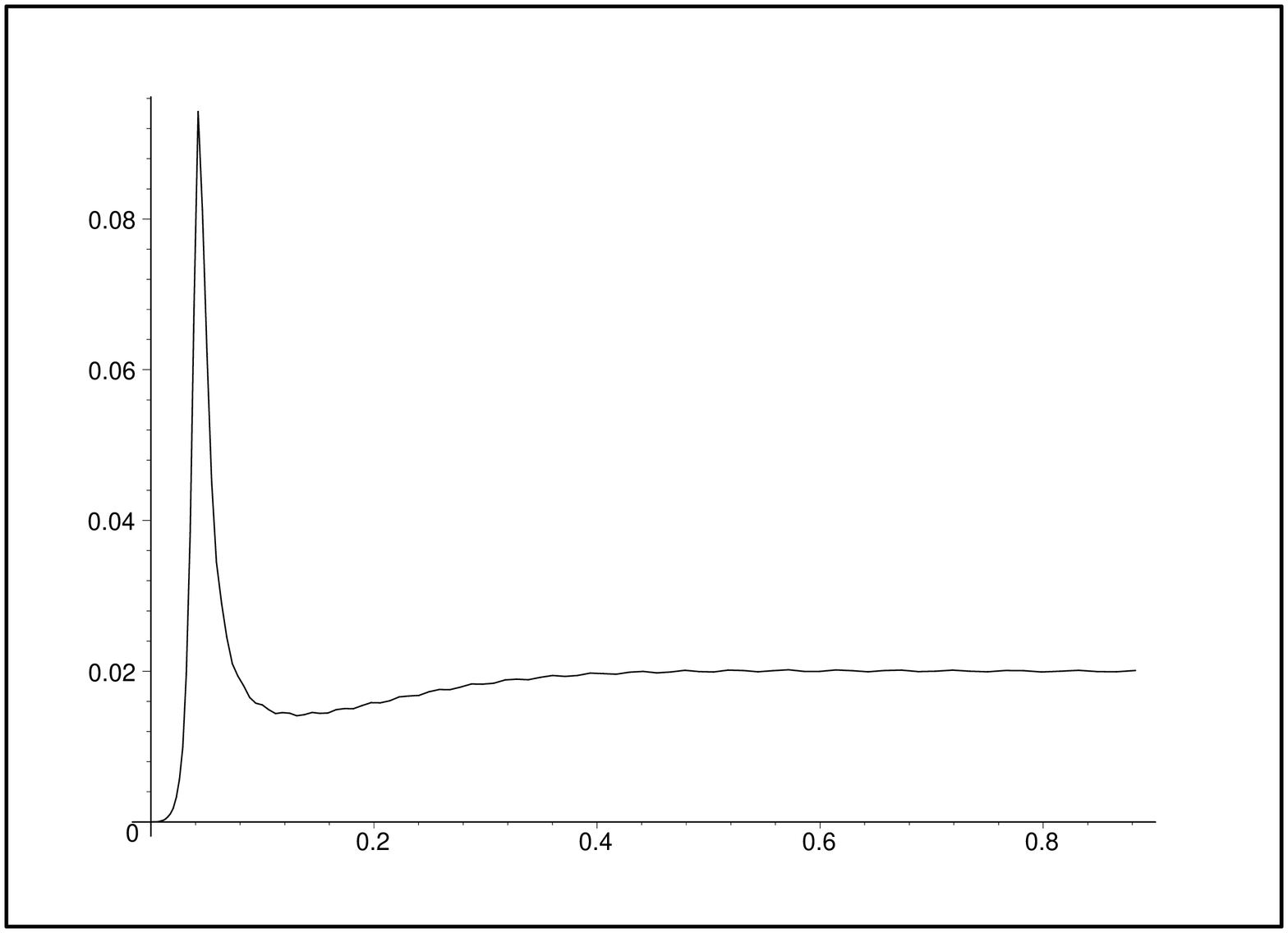}
\includegraphics[angle=-90,width=4.5cm]{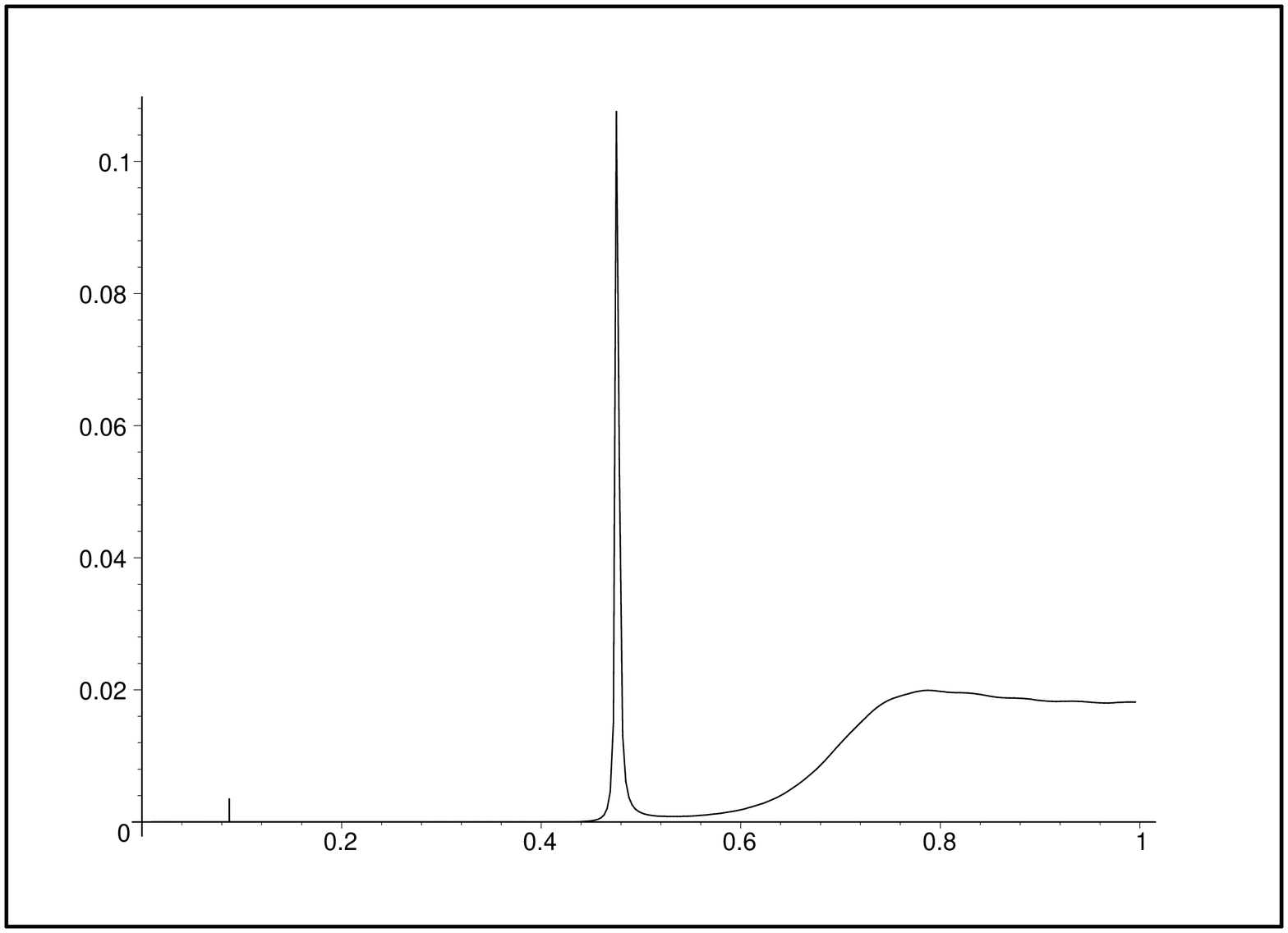}
\includegraphics[angle=-90,width=4.5cm]{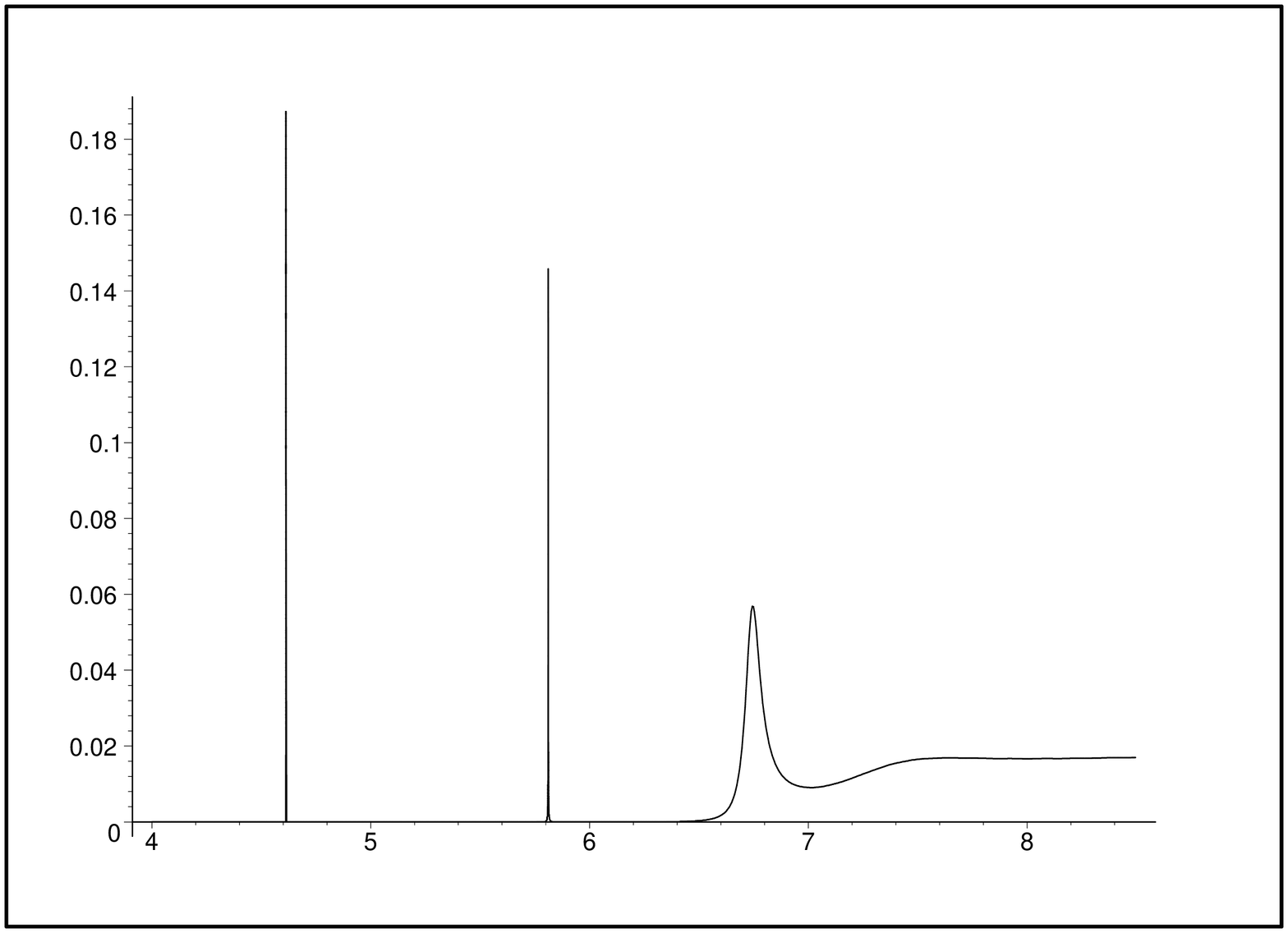}\\
\includegraphics[angle=-90,width=4.5cm]{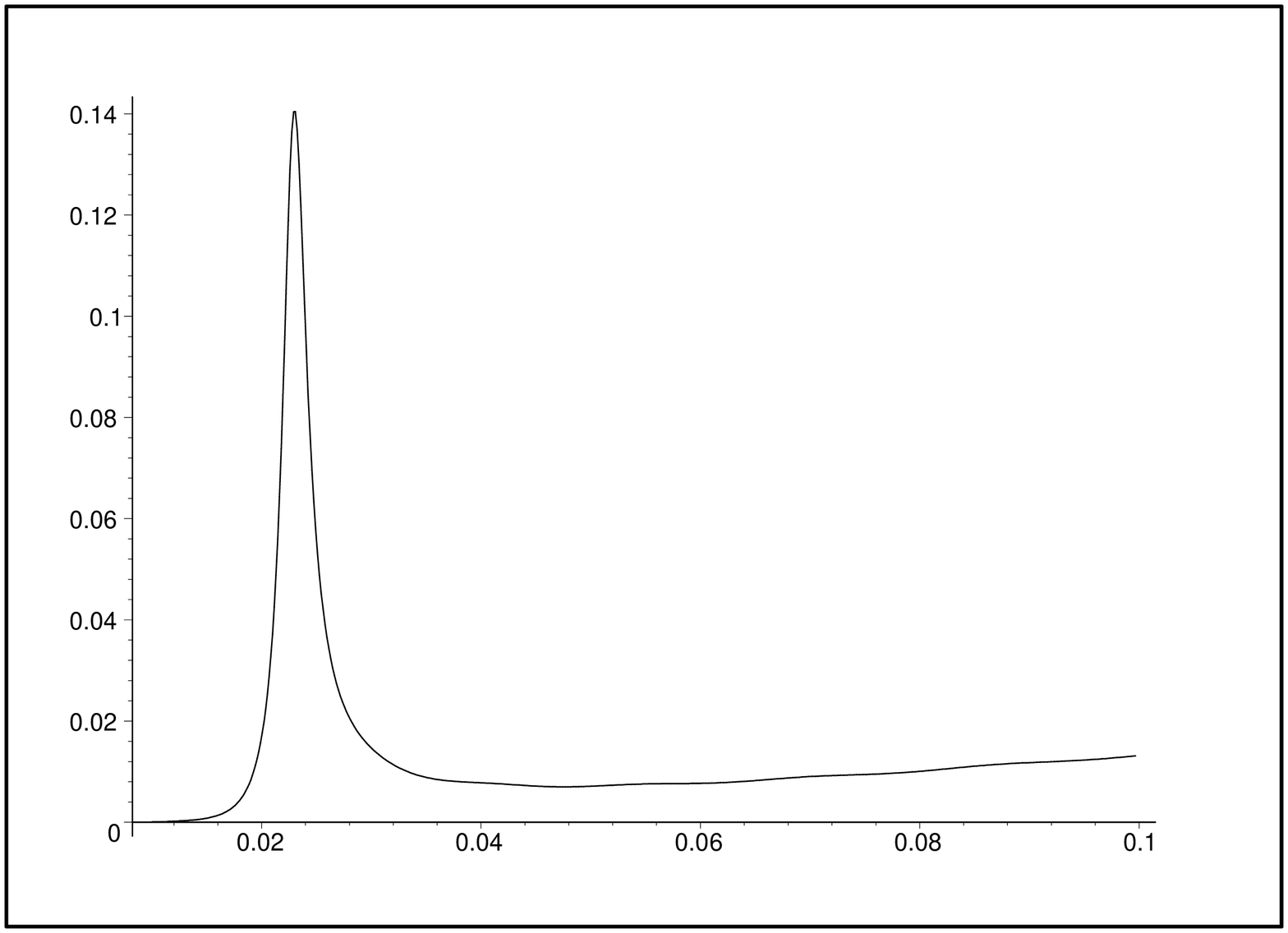}
\includegraphics[angle=-90,width=4.5cm]{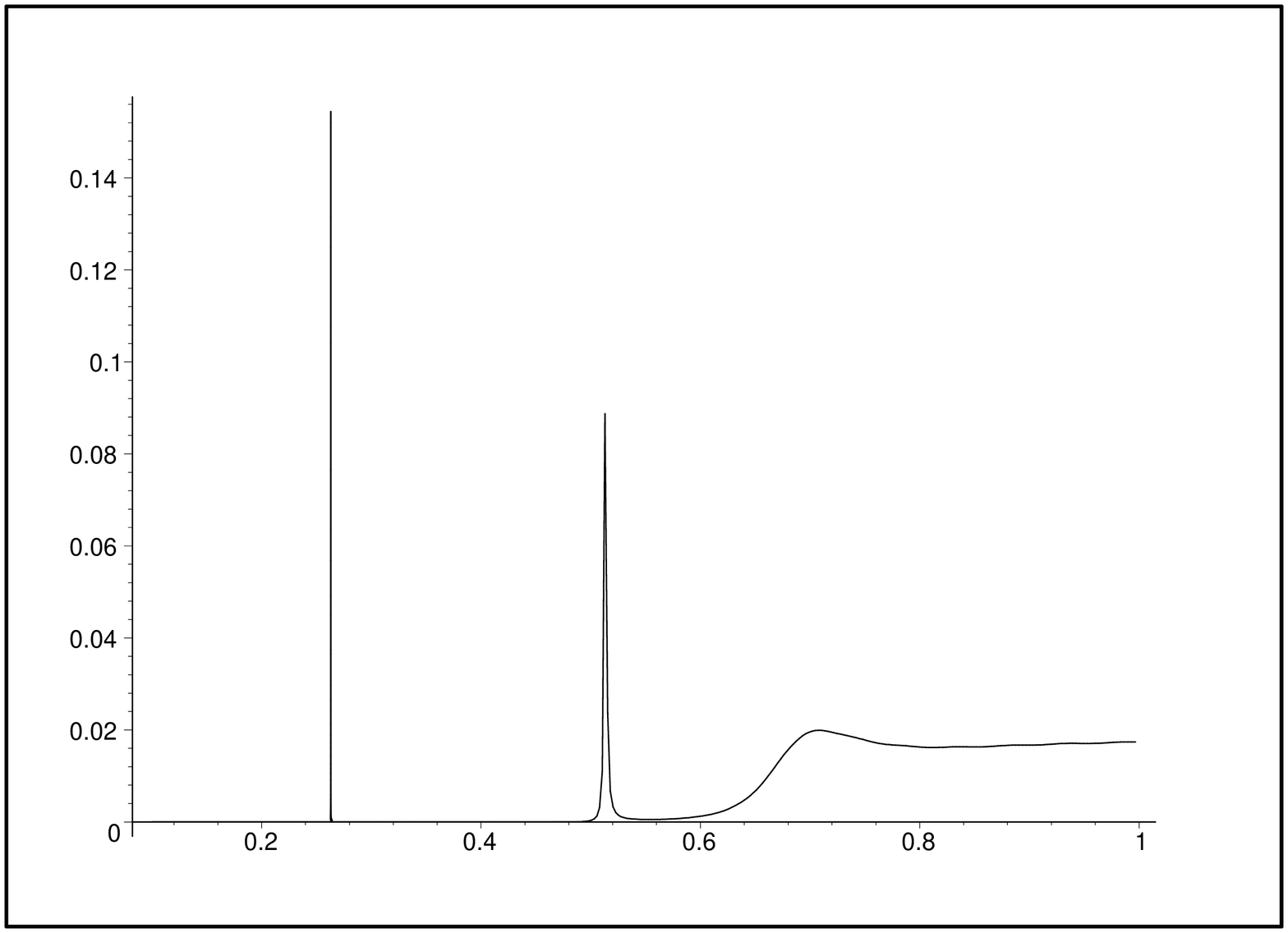}
\includegraphics[angle=-90,width=4.5cm]{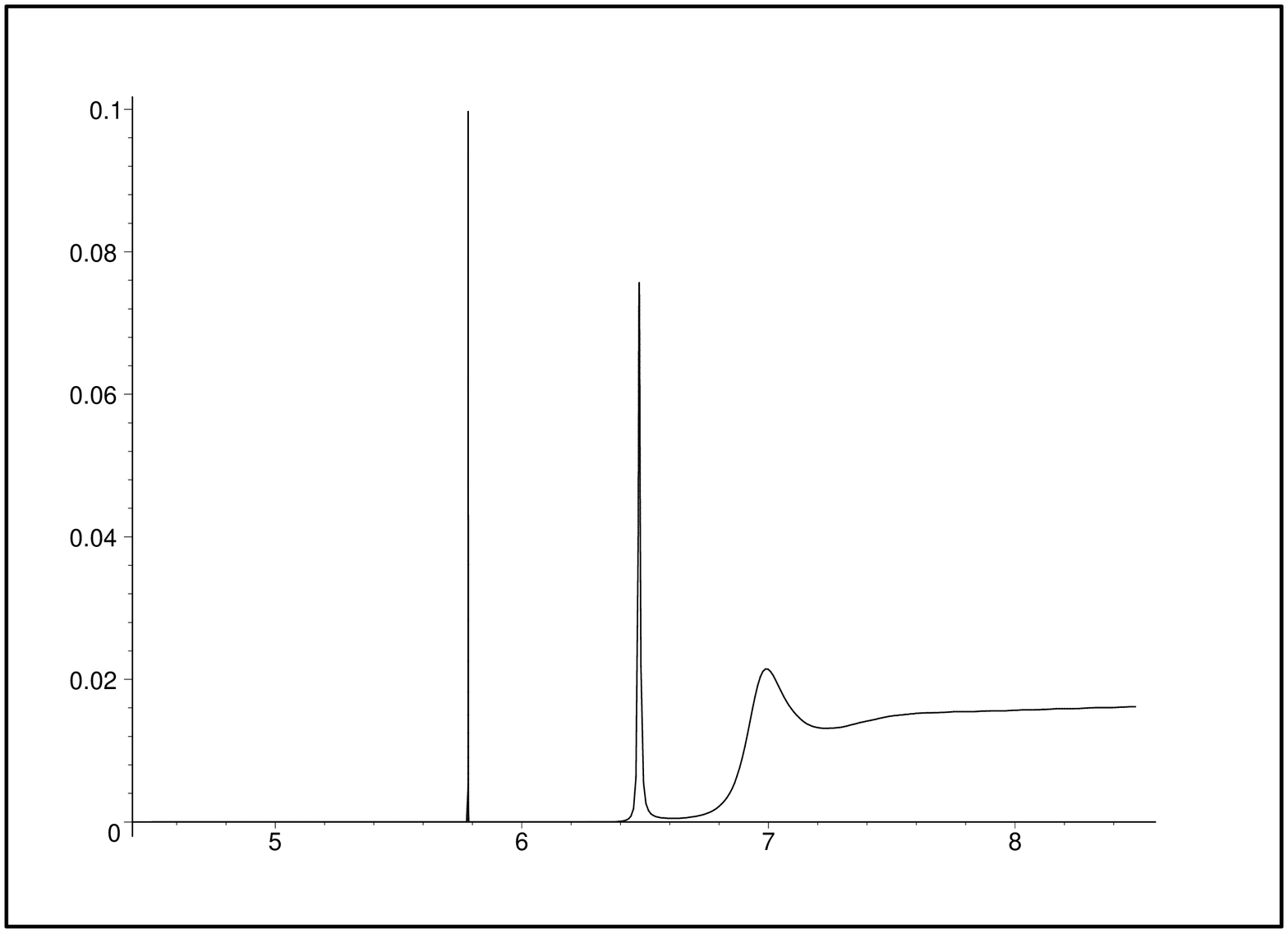}
\caption{\label{probmsq_VR_p3_even} Plots of $|P(0)|^2$ {\it versus} $m^2$ with  $p=3$ (first line), $p=5$ (second line) and $p=7$ (third line). Coupling parameters are $f=0.5$ (left figures), $f=1.1$ (middle figures), and $f=2$ (right figures), for even parity wavefunctions of fermions with right chirality.}
\end{figure}
\begin{figure}
\includegraphics[angle=-90,width=4.5cm]{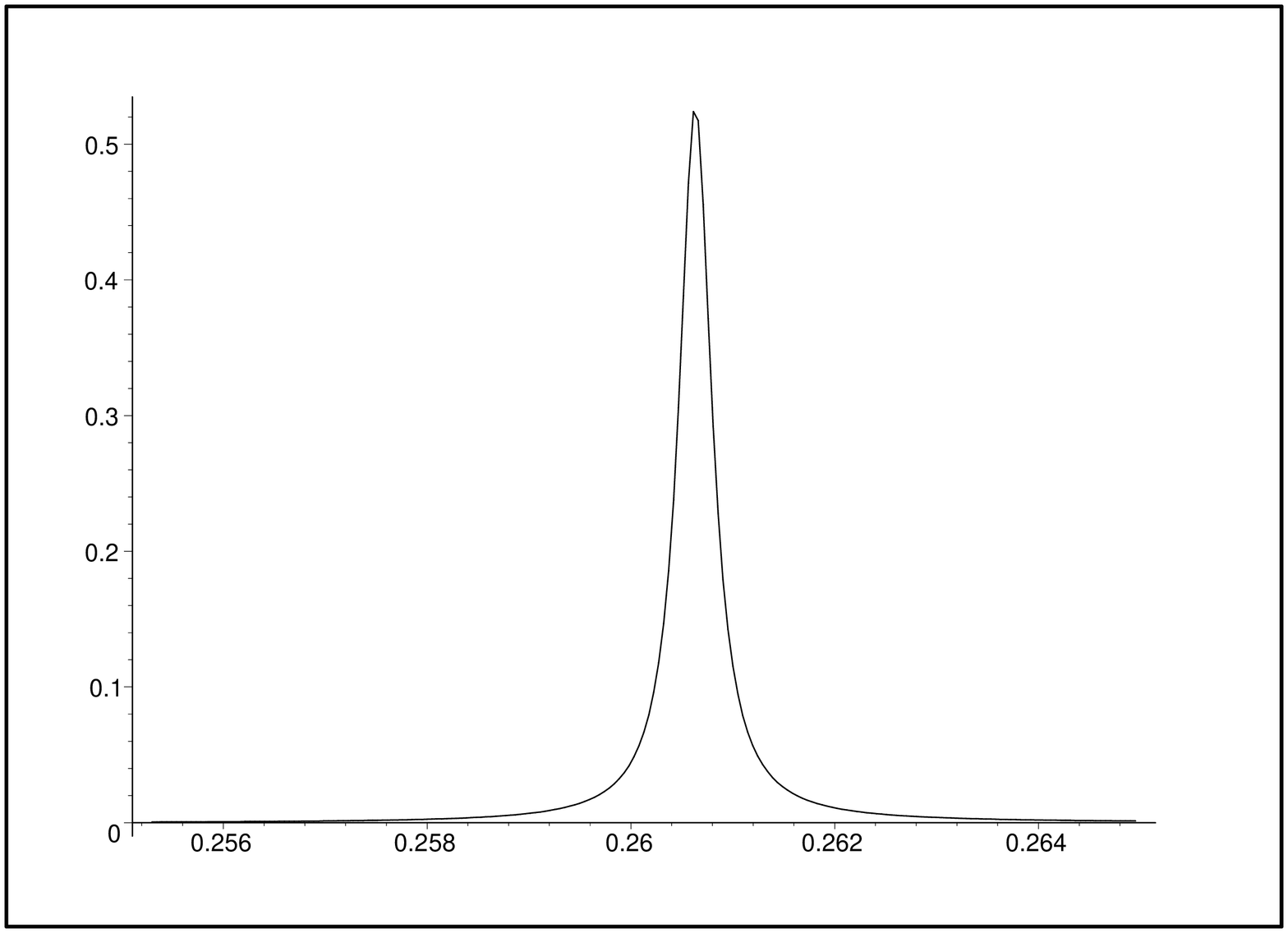}
\includegraphics[angle=-90,width=4.5cm]{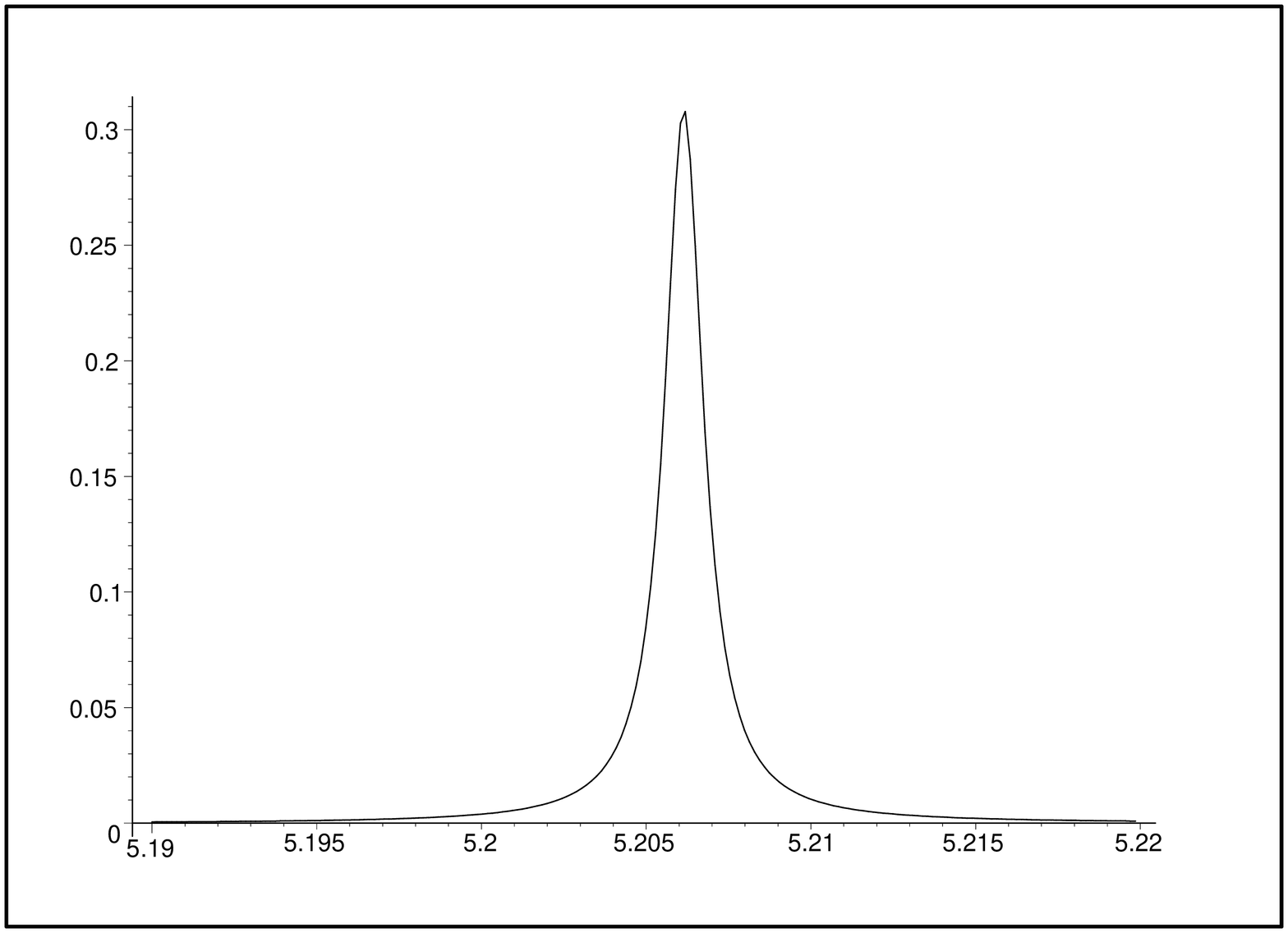}\\
\includegraphics[angle=-90,width=4.5cm]{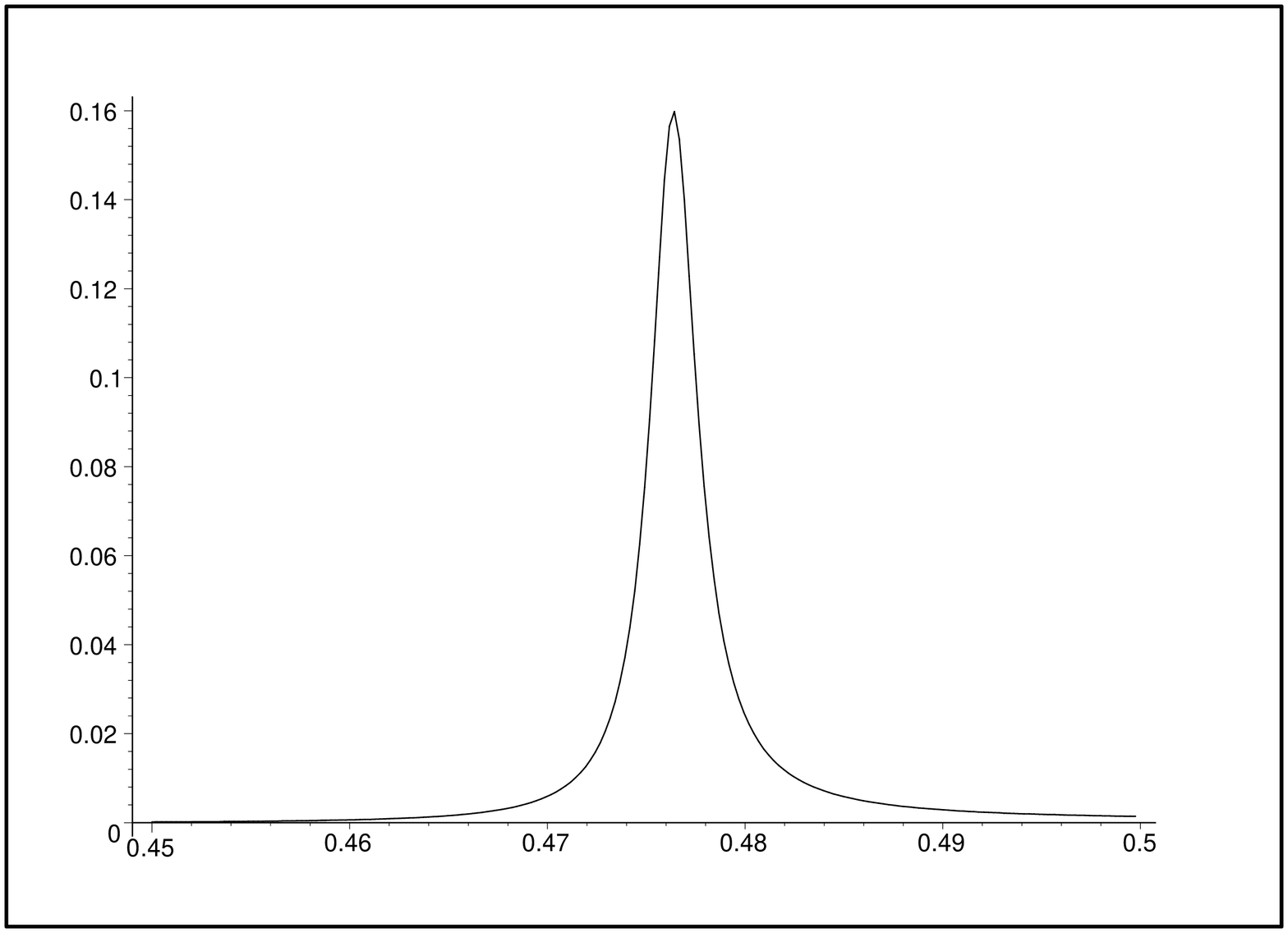}
\includegraphics[angle=-90,width=4.5cm]{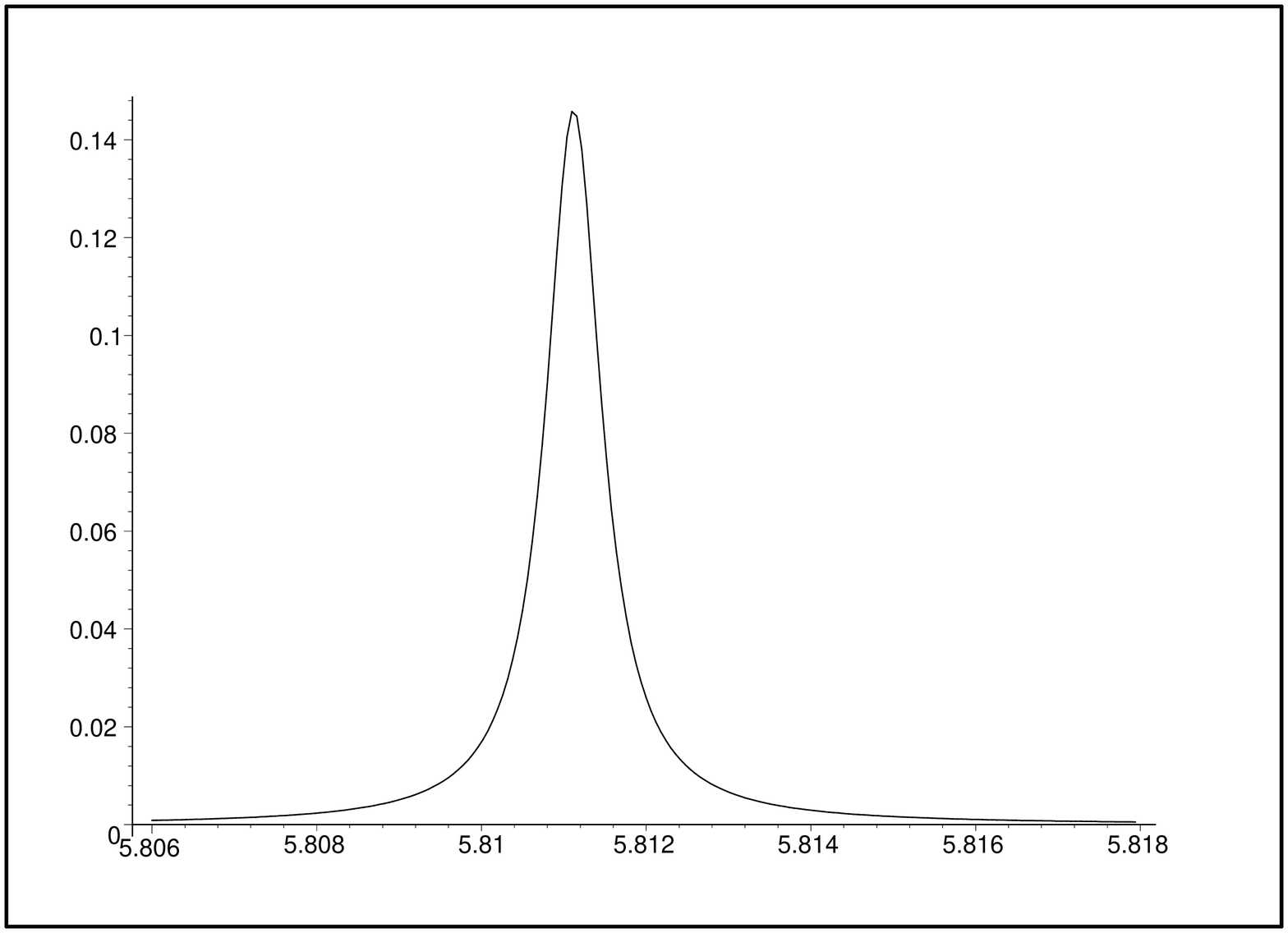}\\
\includegraphics[angle=-90,width=4.5cm]{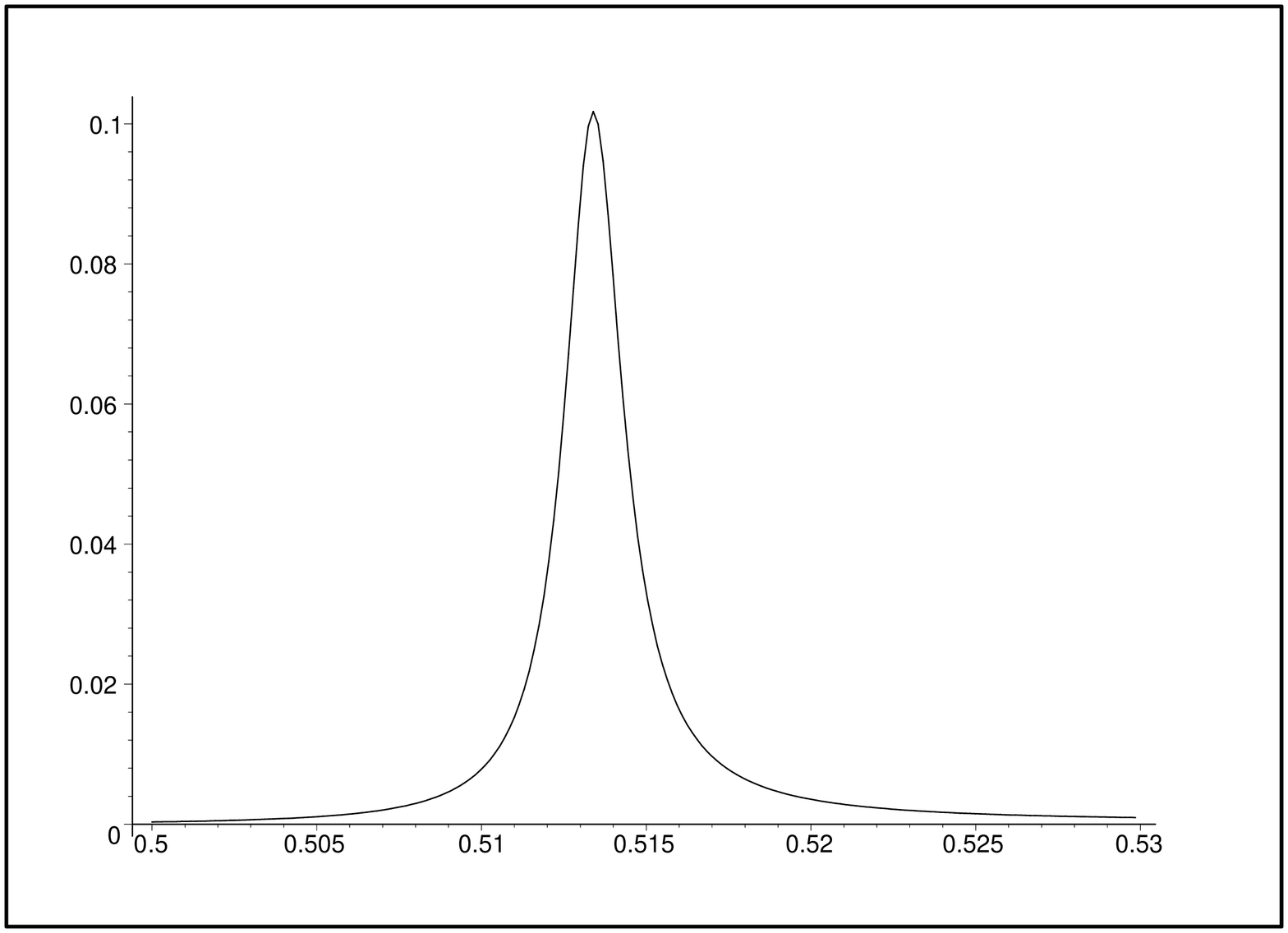}
\includegraphics[angle=-90,width=4.5cm]{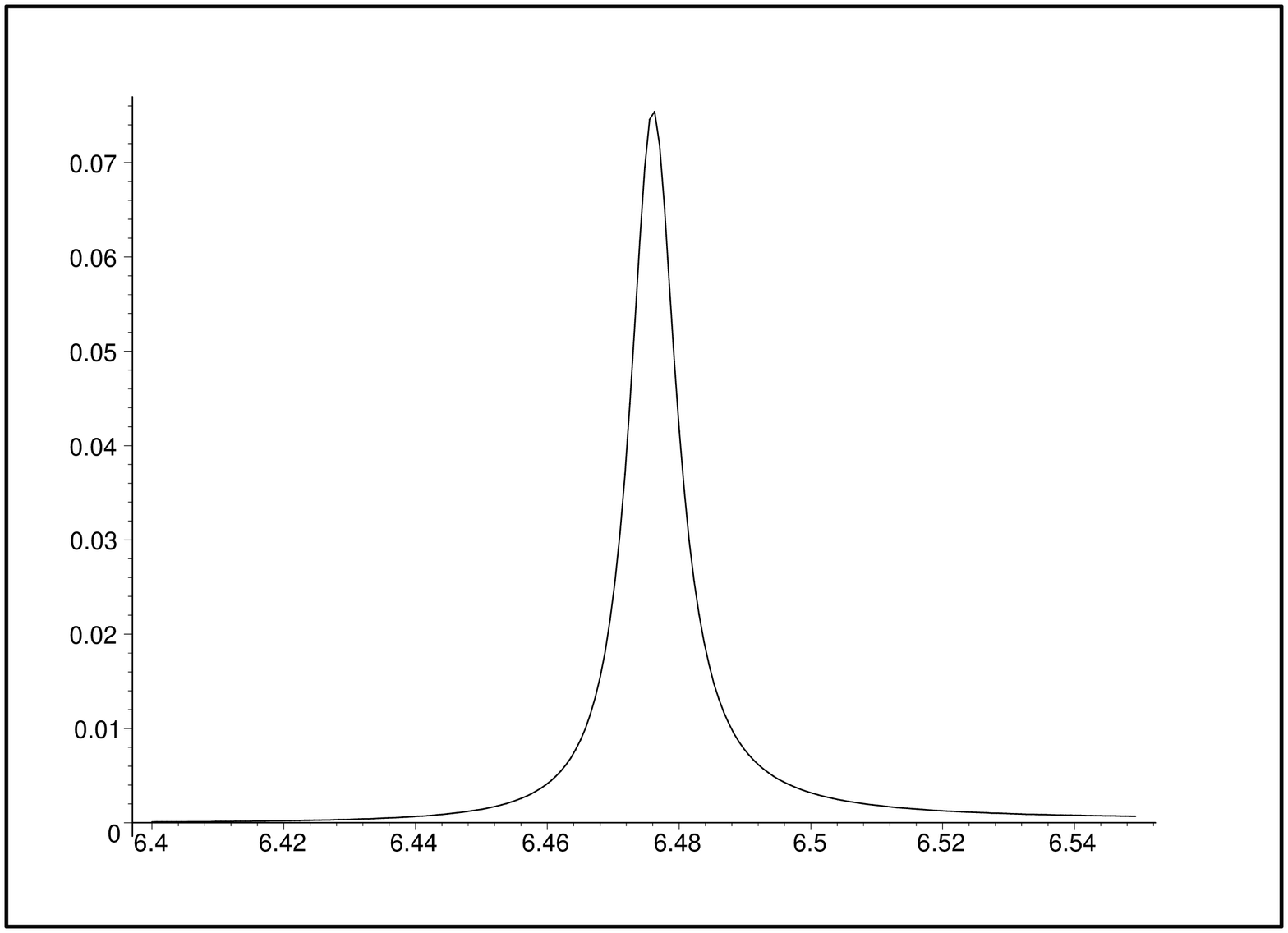}
\caption{\label{probmsq_VR_p3_even_zoom}
Detailed plots of some right chiral resonance peaks for $|P(0)|^2$ {\it versus} $m^2$, corresponding to $f=1.1$ (left) and $f=2$ (right) from Fig. \ref{probmsq_VR_p3_even}.}
\end{figure}
We found resonance peaks for even parity wavefunctions and analyzed the influence of the parameter $p$ and coupling constant $f$.

Some results for right chiral fermions are shown in Fig. \ref{probmsq_VR_p3_even} for $p=3, 5$ and $7$ and three values of $f$. Fig. \ref{probmsq_VR_p3_even_zoom} shows in detail some thin resonant peaks corresponding to Fig. \ref{probmsq_VR_p3_even}, with the corresponding values of $m^2$ identified in Table I. For $p=3$ and $f=0.5$ we see that there is no resonance, as the width at half maximum $\Delta m$ of the peak is bigger than the mass $m$ corresponding to the peak; for $f=1.1$ and $f=2$ there appears one and three resonances, respectively. This agrees with our expectation that larger values of $f$ favor the presence of resonances. Note that in general, for fixed $p$, larger values of $f$ lead to larger number of peaks. Also, for fixed $f$, larger values of $p$ corresponds to the thicker peaks with almost the same mass. The first peak is the thinnest, with corresponding larger lifetime. The increasing of the parameter $p$ tends to increase the mass of the first resonance (this is more evident for larger values of $f$), whereas the increasing of $f$ turns richer the spectrum, with more resonances.
{\tiny
\begin{table}[htbp]
        \begin{tabular}
        {|l |c | c|r|}\hline {Right} & $f=0.5$ & $f=1.1$ & $f=2$  \\
        \hline $p=3$
 & absent & 0.26061 & 2.76451953; 5.2062; 6.94  \\
        \hline $p=5$ & 0.042 & 0.0873678916; 0.4764 & 4.6115419;5.8111; 6.746  \\
        \hline $p=7$
 & 0.0230 & 0.26301844; 0.5134 & 4.99069057;5.782825;6.476;6.99106 \\
        \hline
        \end{tabular}
               \caption{\it First resonance peaks, even parity modes, right chirality. The table shows the corresponding values of $m^2$.}
   \end{table}}

\begin{figure}
\includegraphics[angle=-90,width=4.5cm]{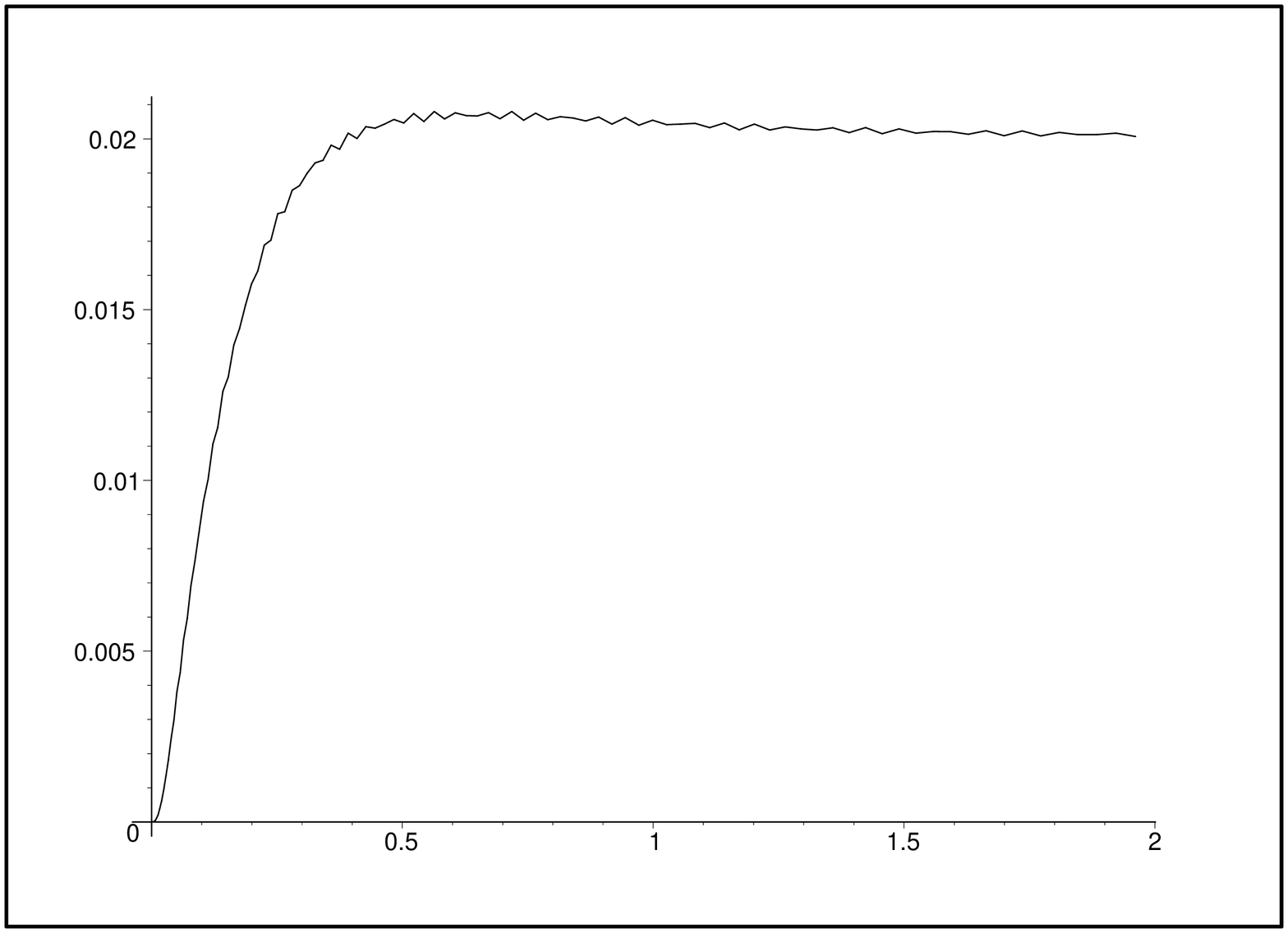}
\includegraphics[angle=-90,width=4.5cm]{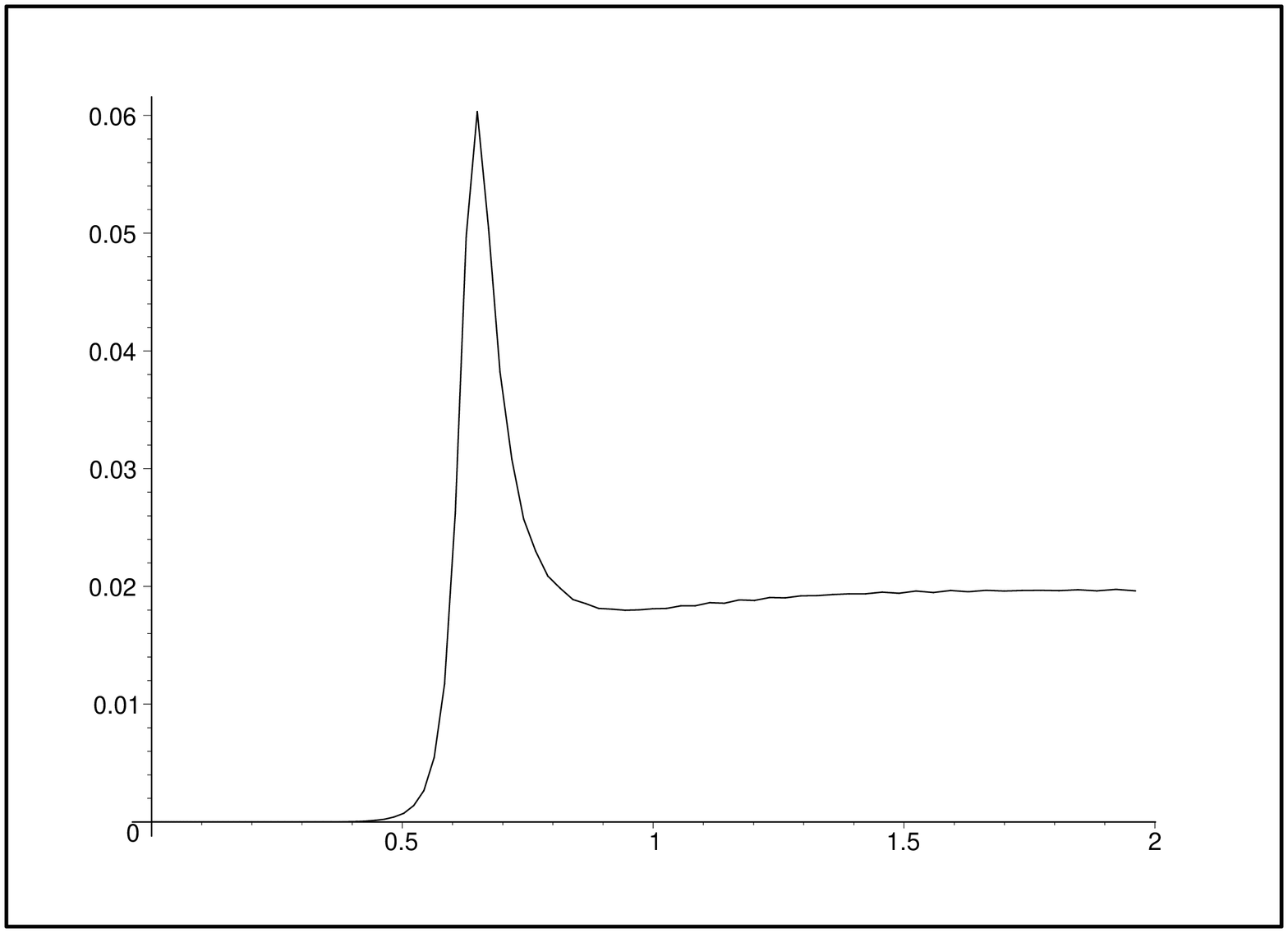}
\includegraphics[angle=-90,width=4.5cm]{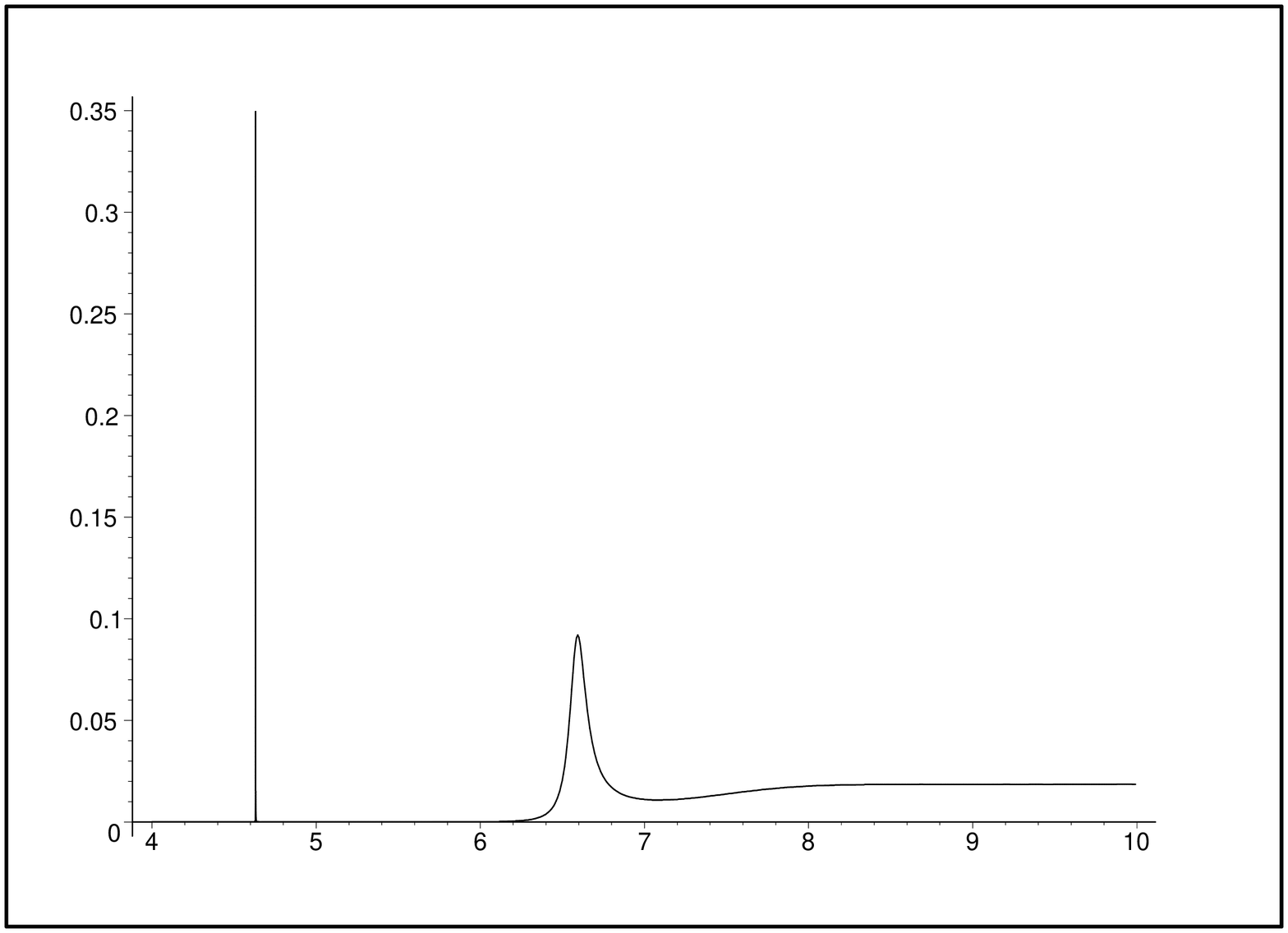}
\includegraphics[angle=-90,width=4.5cm]{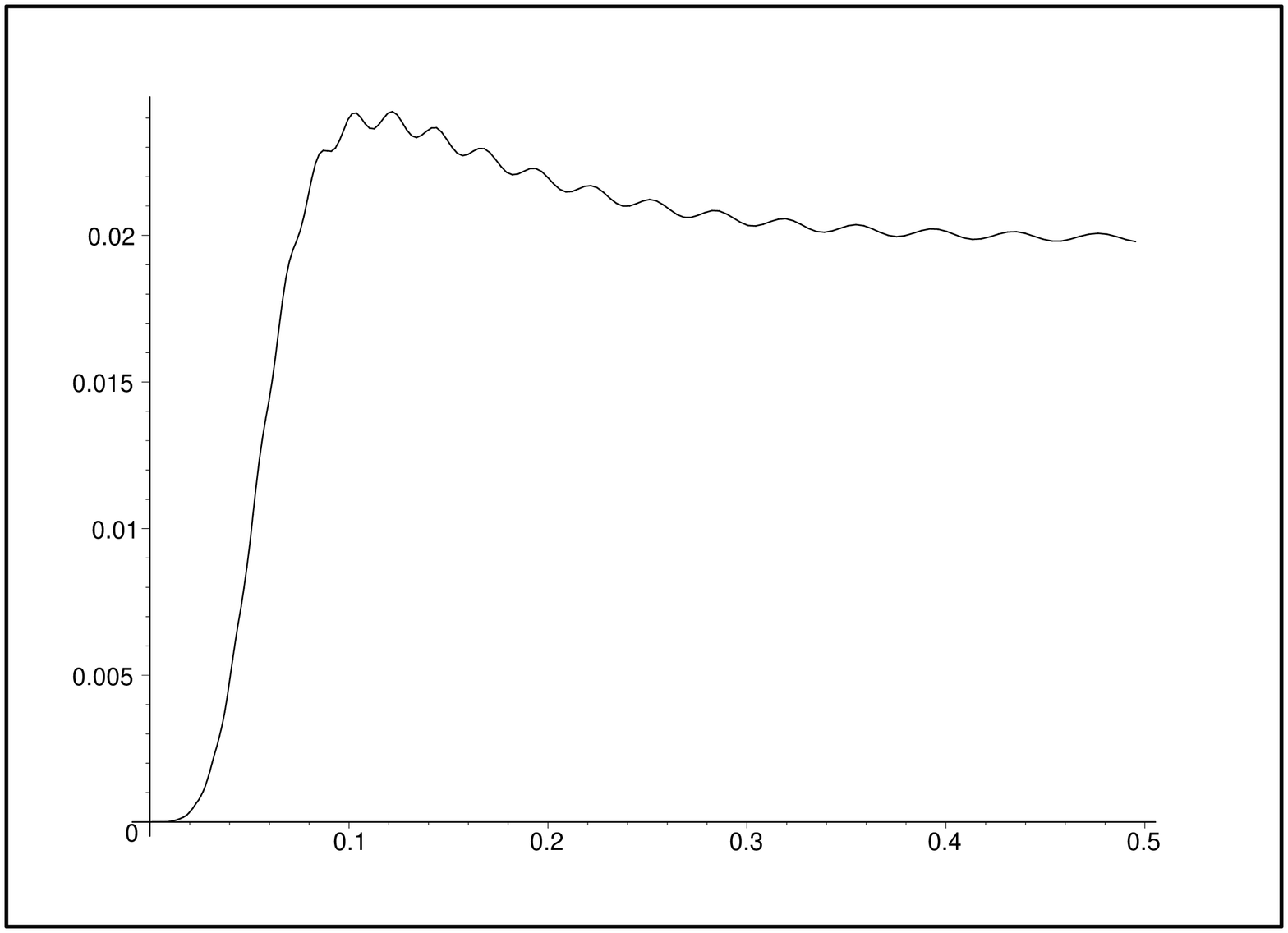}
\includegraphics[angle=-90,width=4.5cm]{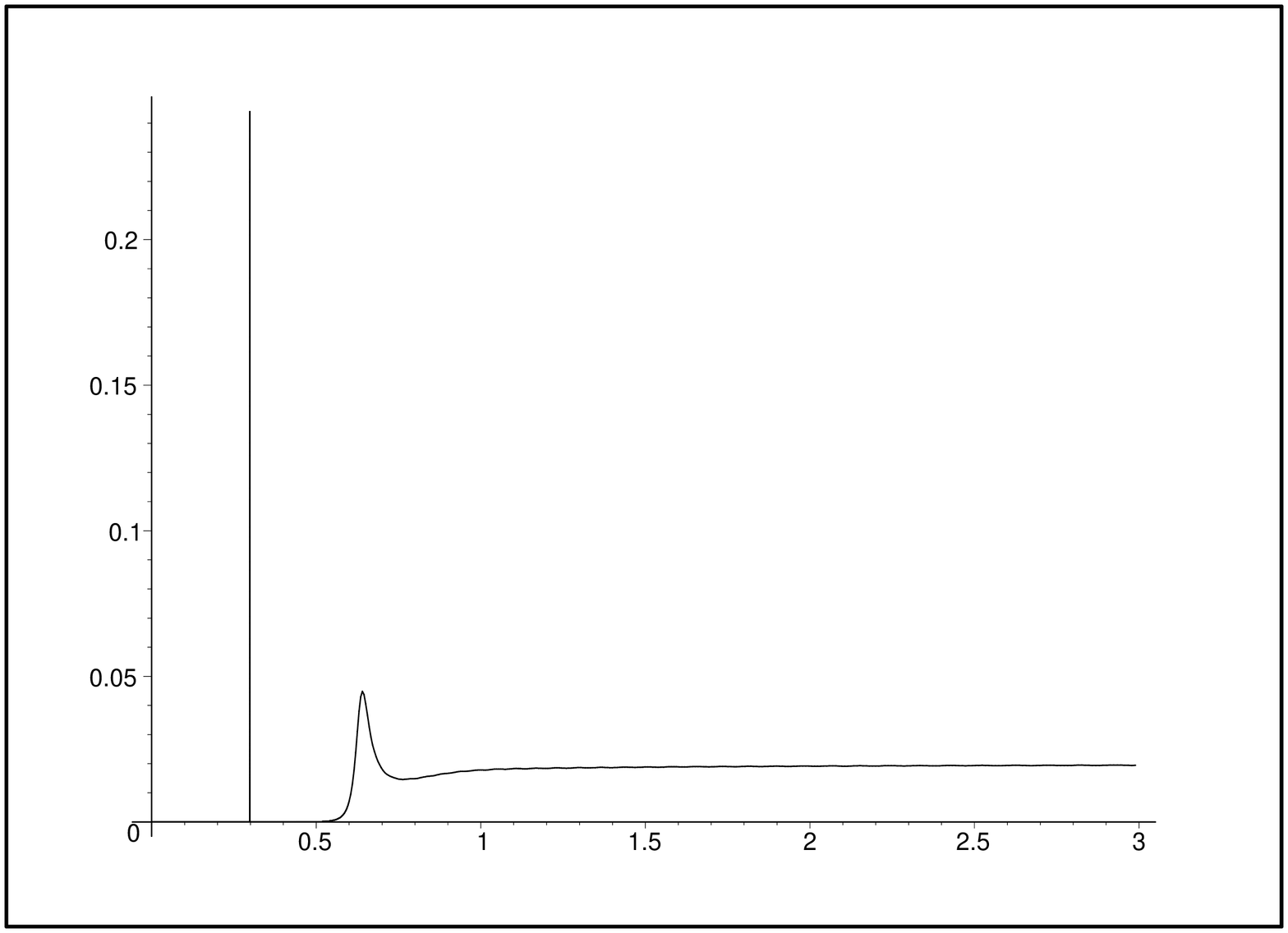}
\includegraphics[angle=-90,width=4.5cm]{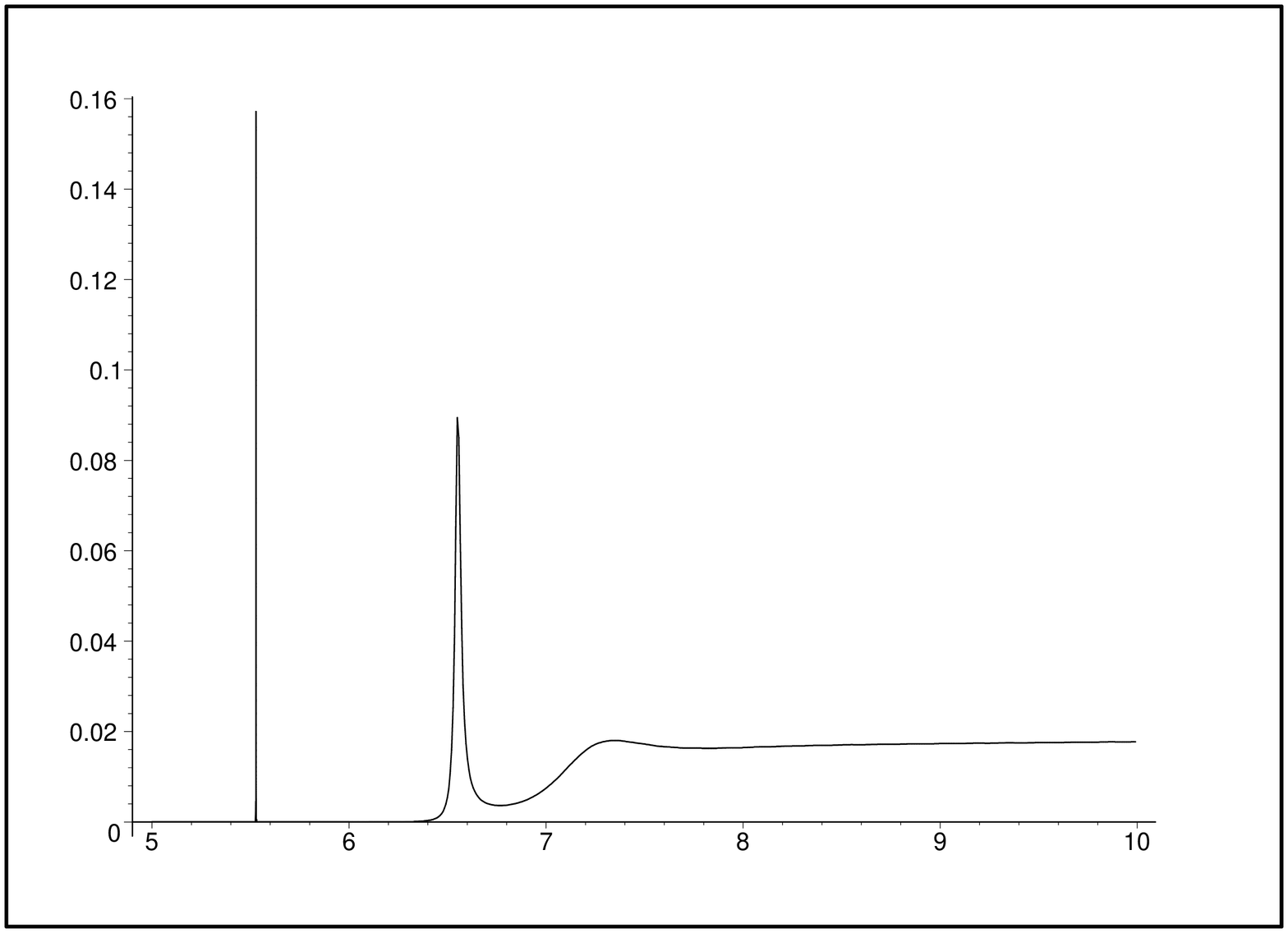}
\includegraphics[angle=-90,width=4.5cm]{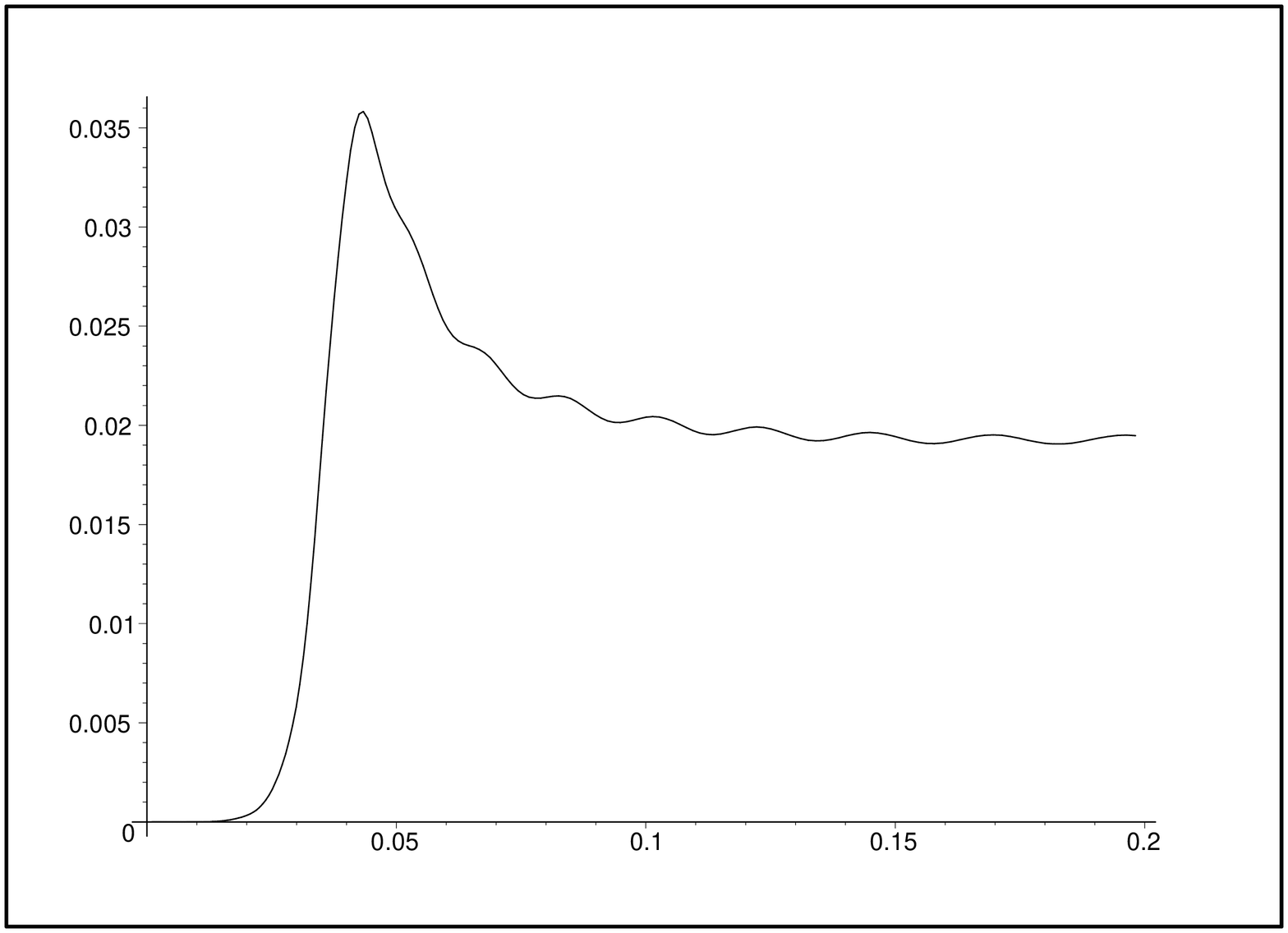}
\includegraphics[angle=-90,width=4.5cm]{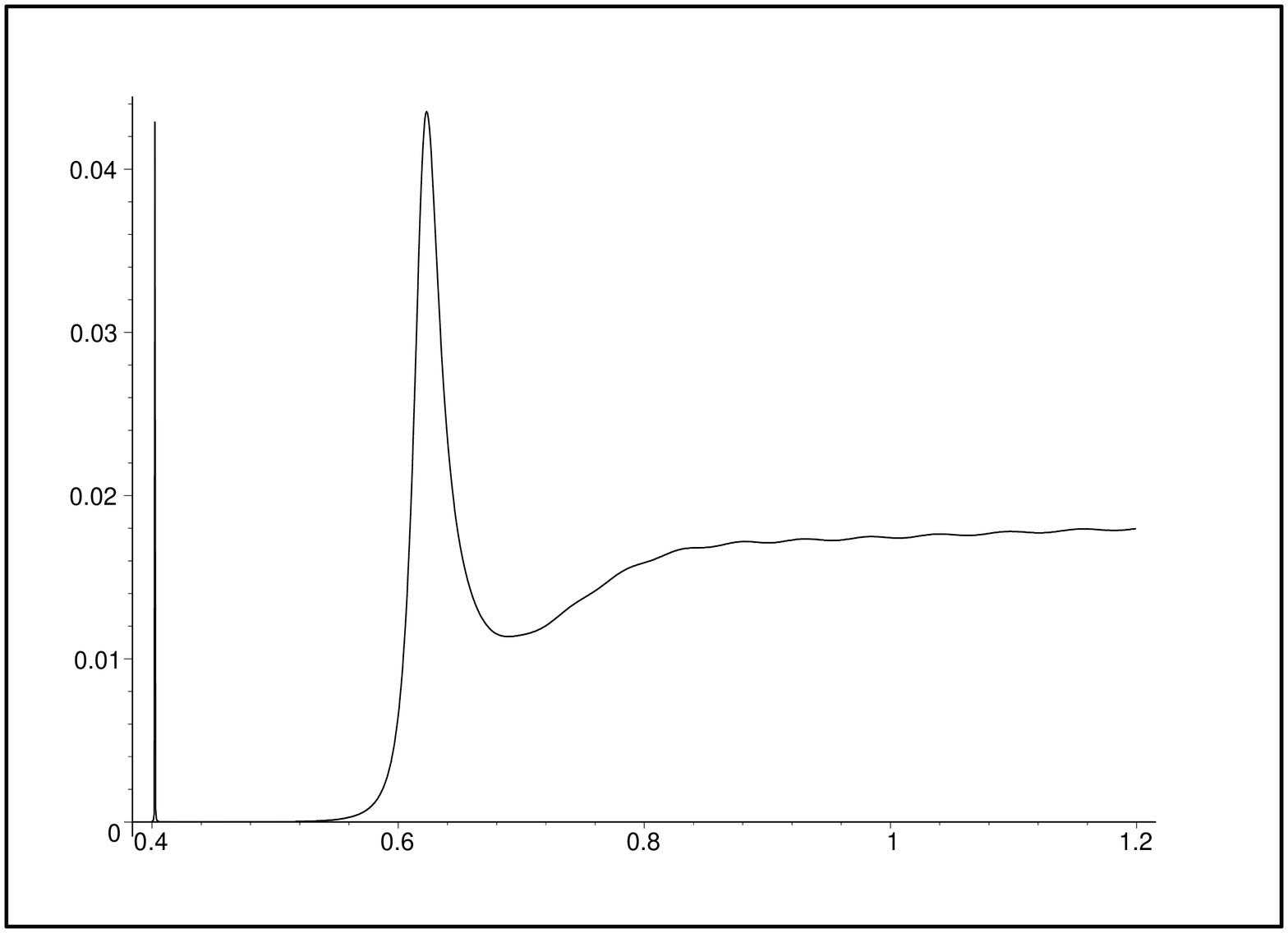}
\includegraphics[angle=-90,width=4.5cm]{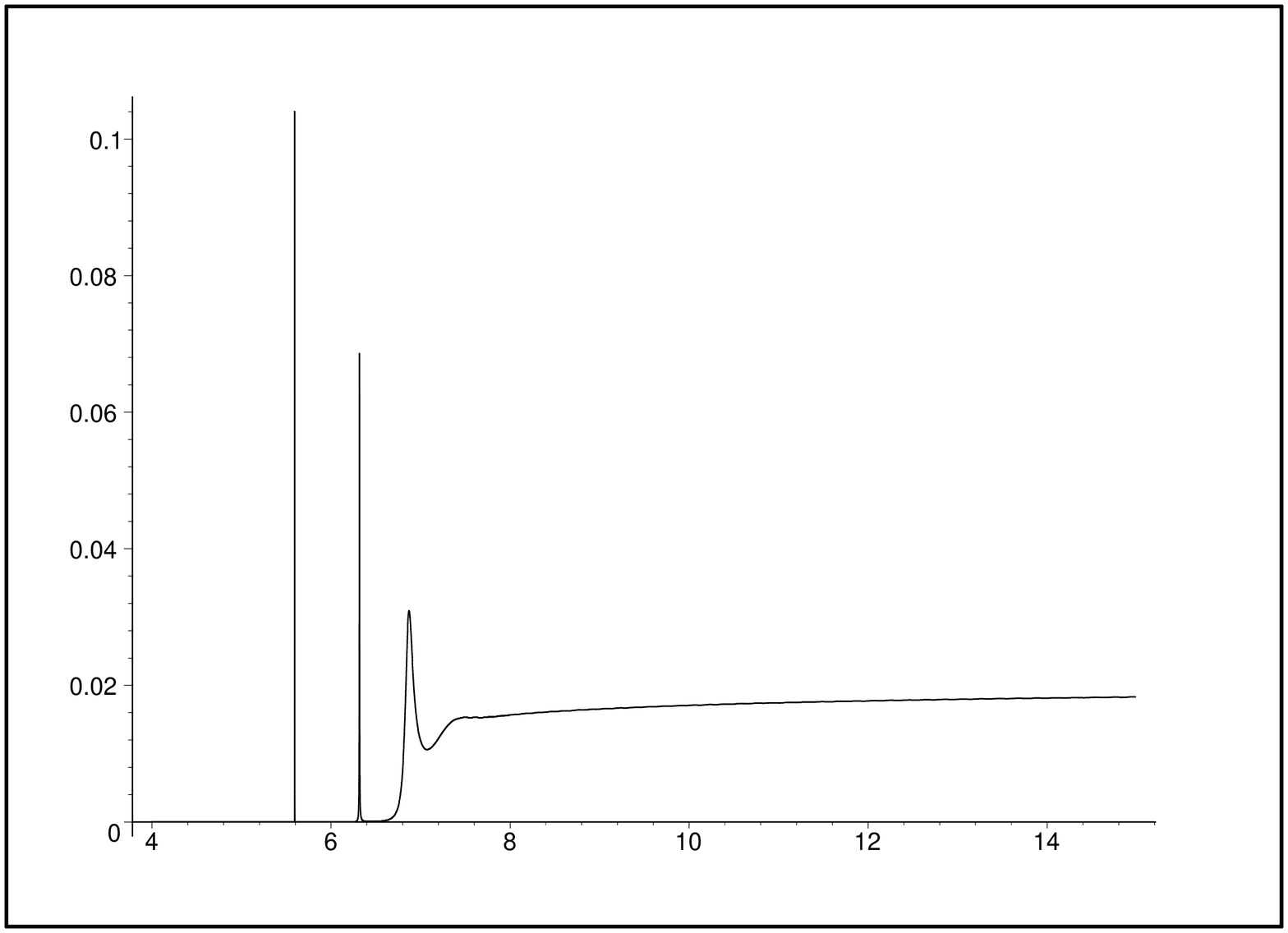}
\caption{\label{probmsq_VL_p3_even} Plots of $|P(0)|^2$ {\it versus} $m^2$ with  $p=3$ (first line), $p=5$ (second line) and $p=7$ (third line). Coupling parameters are $f=0.5$ (left figures), $f=1.1$ (middle figures), and $f=2$ (right figures), for even parity wavefunctions of fermions with left chirality.}
\end{figure}

\begin{figure}
\includegraphics[angle=-90,width=4.5cm]{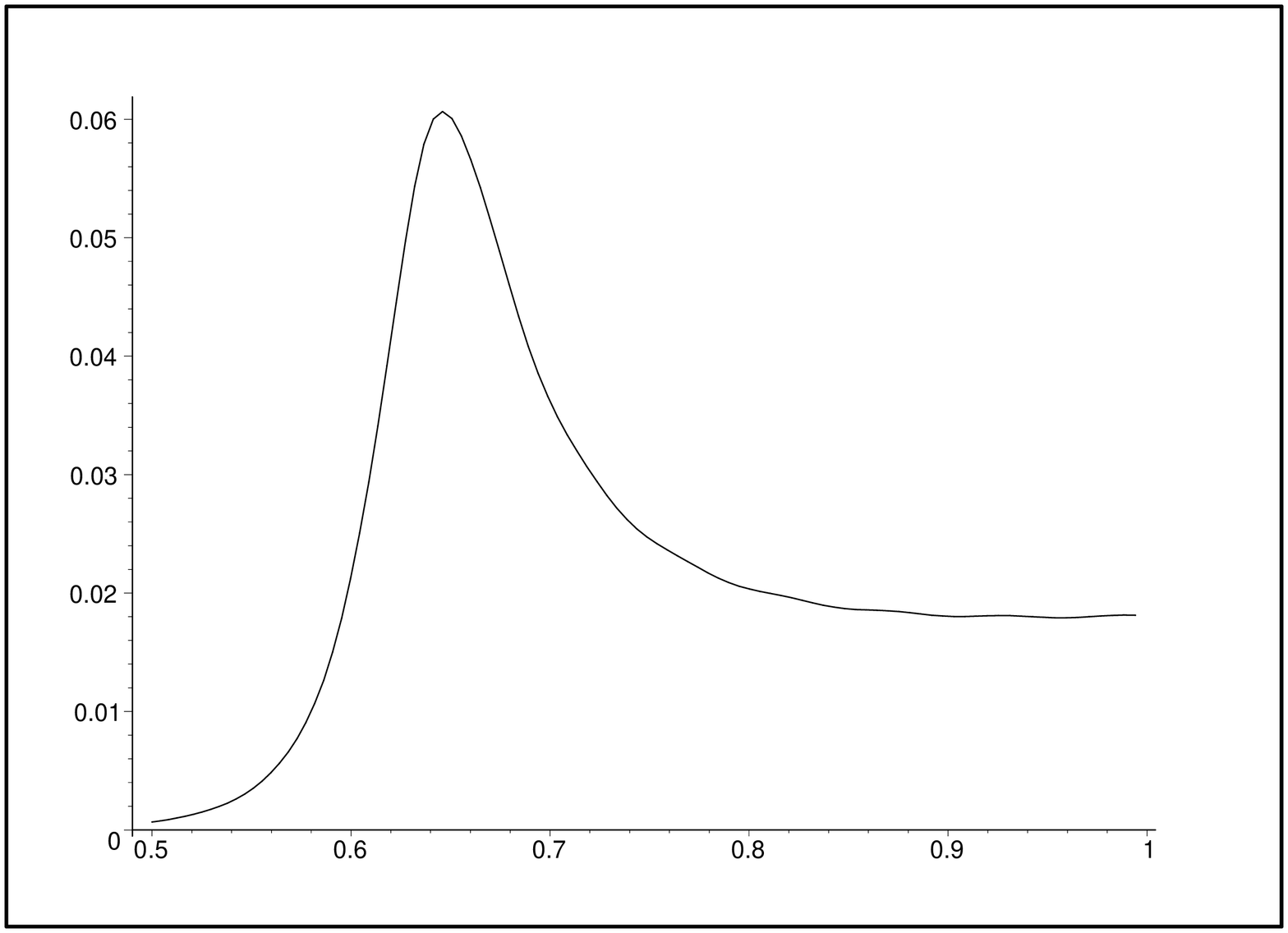}
\includegraphics[angle=-90,width=4.5cm]{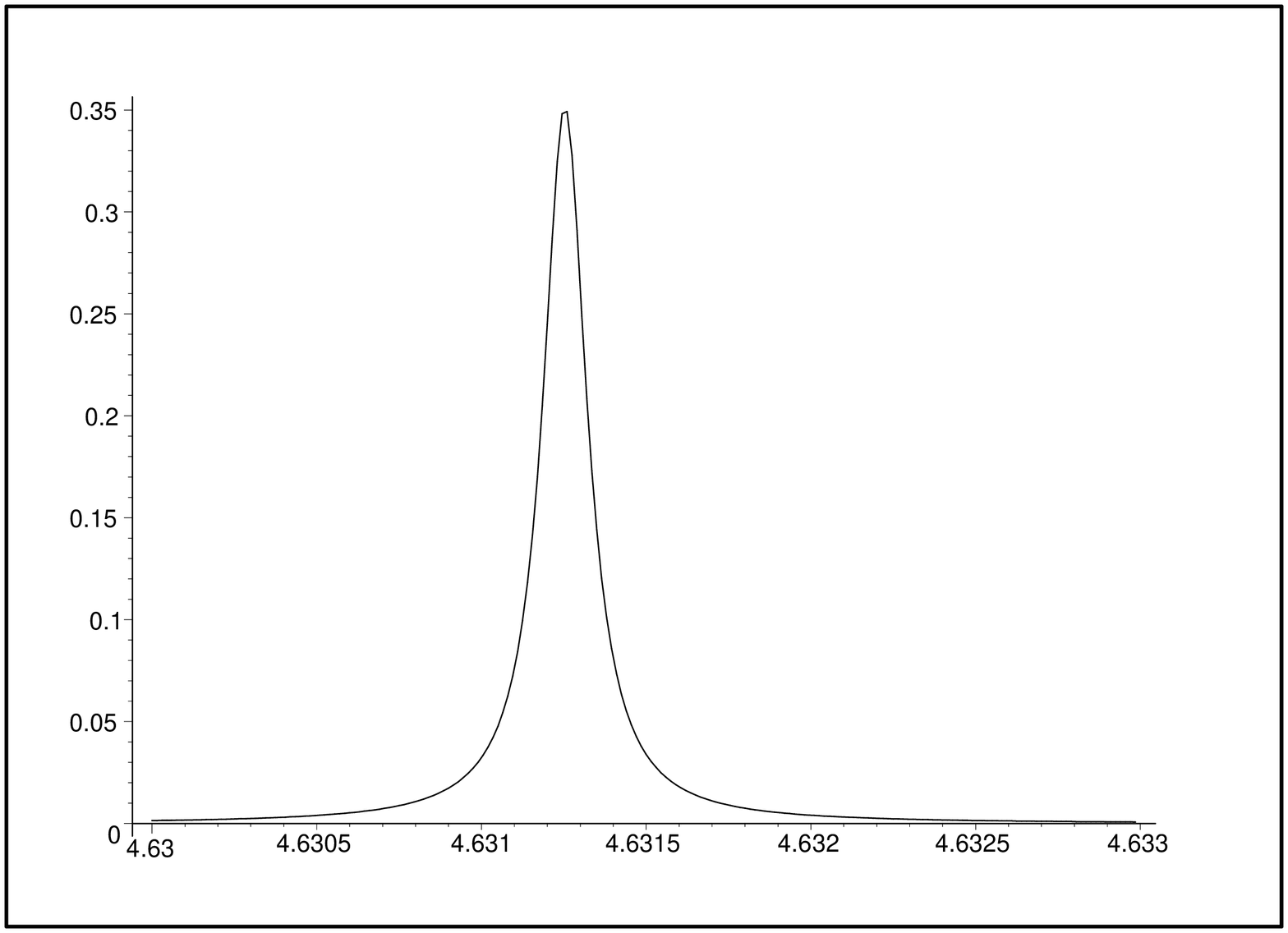}\\
\includegraphics[angle=-90,width=4.5cm]{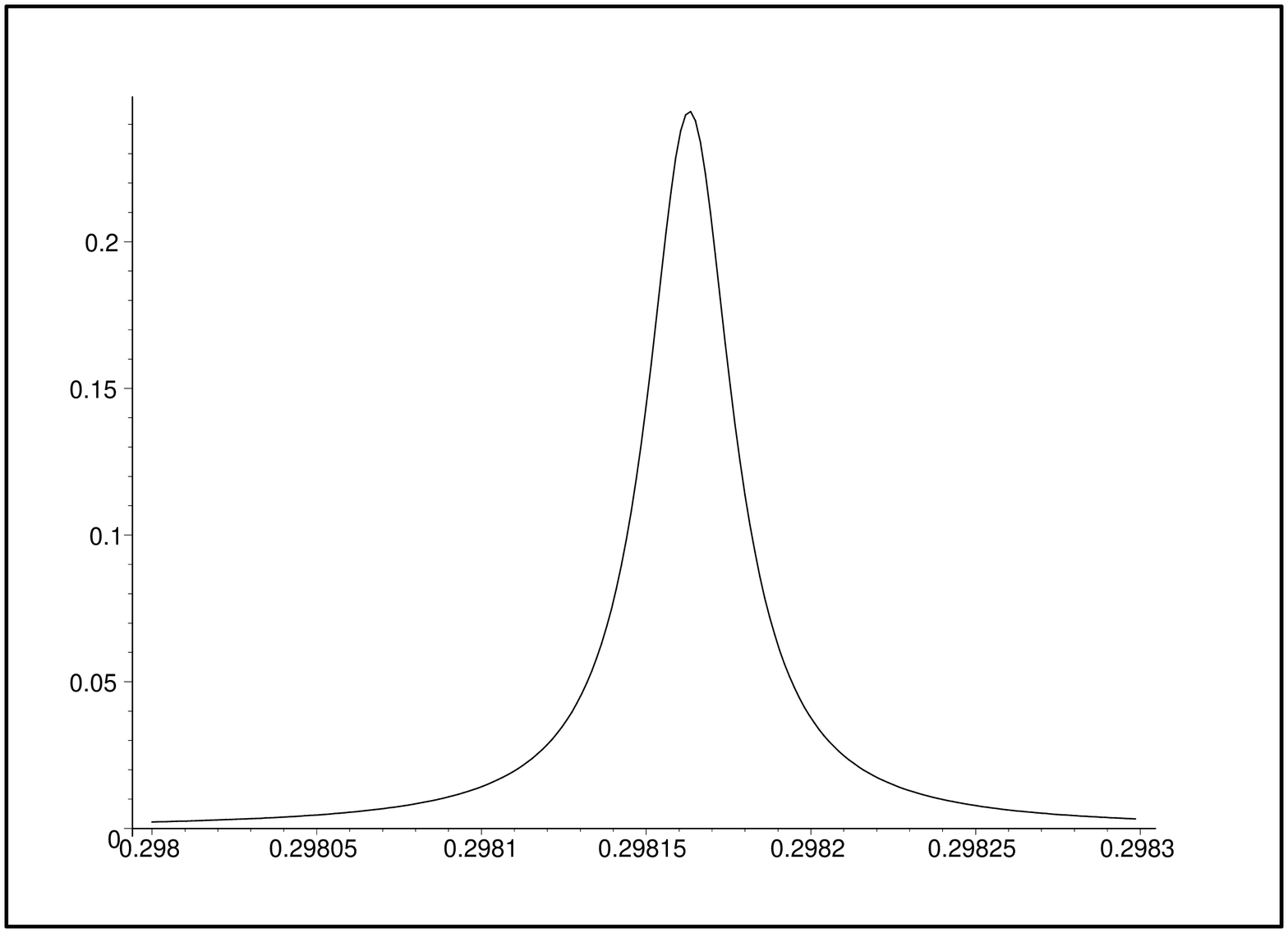}
\includegraphics[angle=-90,width=4.5cm]{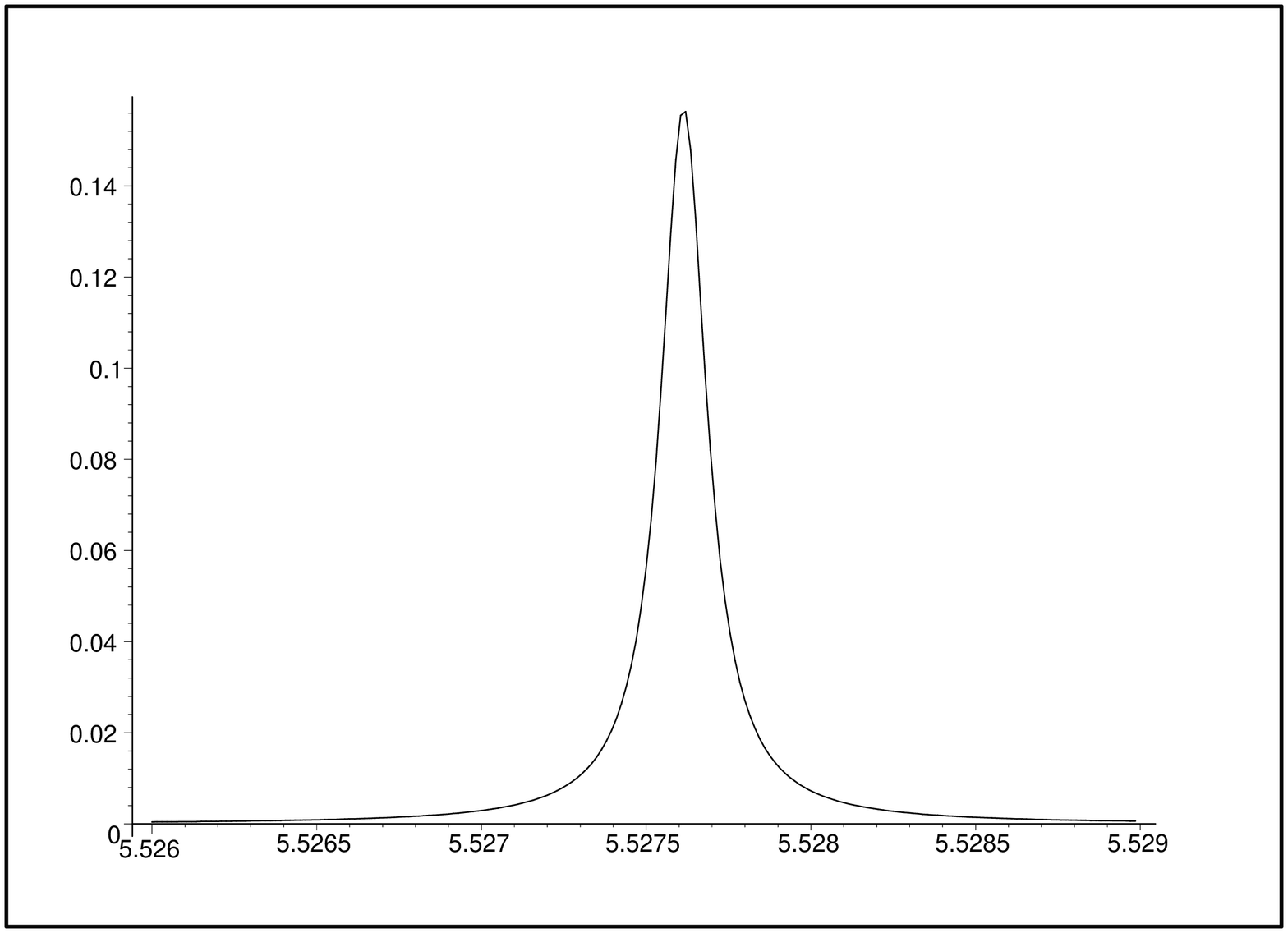}\\
\includegraphics[angle=-90,width=4.5cm]{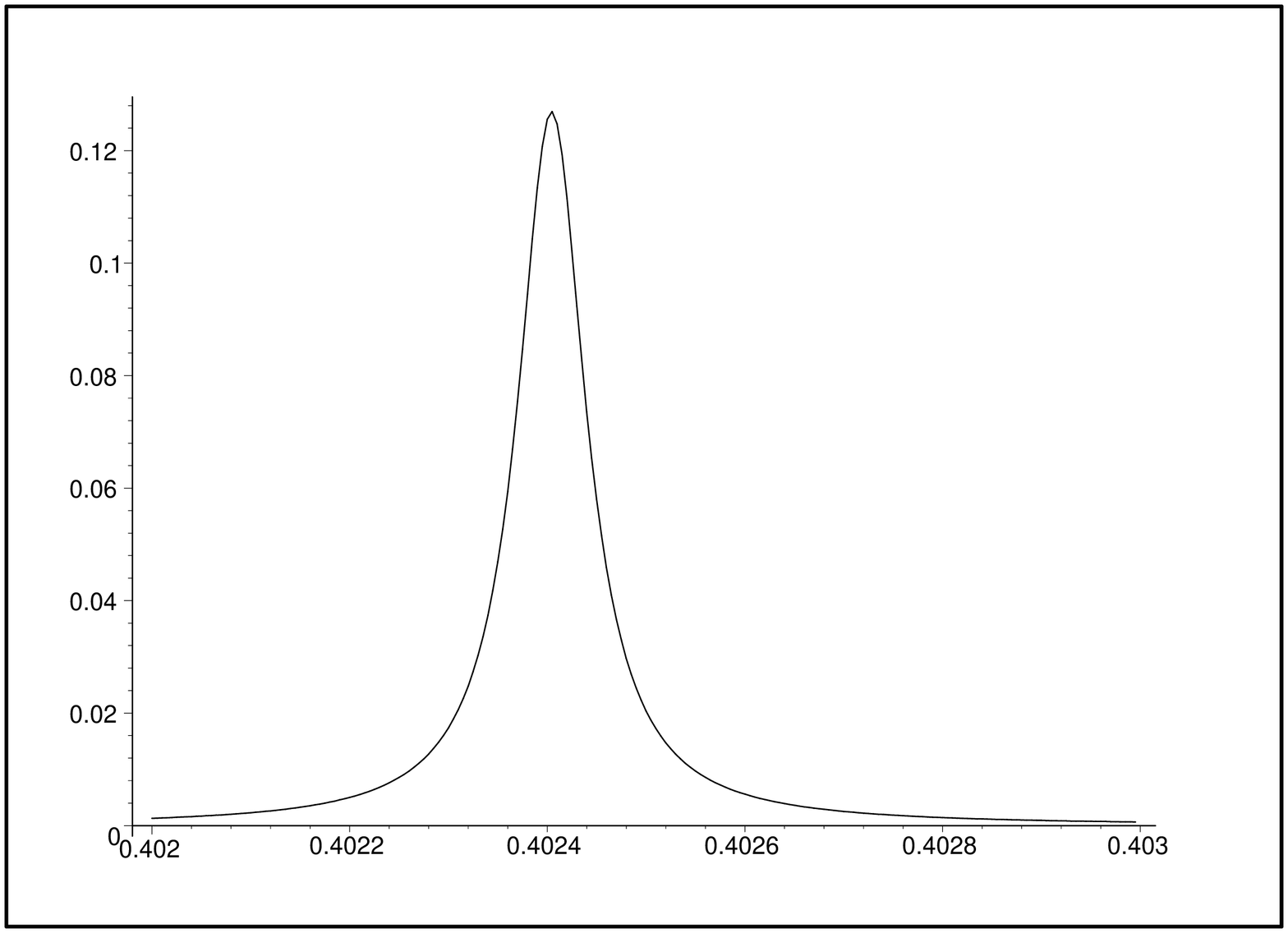}
\includegraphics[angle=-90,width=4.5cm]{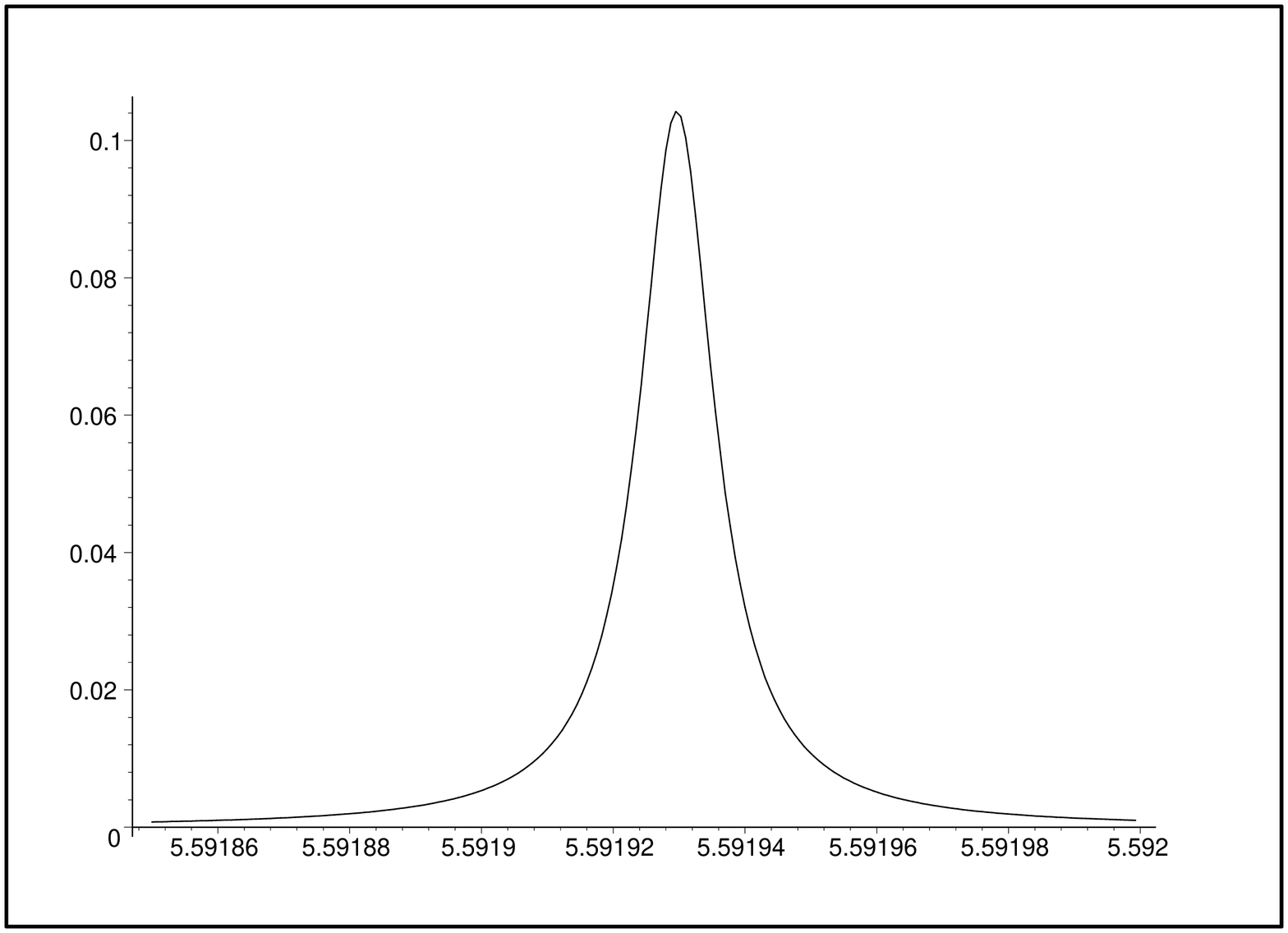}
    \caption{\label{probmsq_VL_p3_even_zoom} Detailed plots of some left chiral resonance peaks for $|P(0)|^2$ {\it versus} $m^2$, corresponding to $f=1.1$ (left) and $f=2$ (right) from Fig. \ref{probmsq_VL_p3_even}.}
\end{figure}

Next we analyzed left chiral fermions. Fig. \ref{probmsq_VL_p3_even} shows the resonance peaks for $p=3,5,7$ with same values of $f$ used previously, but now for left chirality. Fig. \ref{probmsq_VL_p3_even_zoom} shows in detail some thin resonant peaks corresponding to Fig. \ref{probmsq_VL_p3_even}, with the corresponding values of $m^2$ identified in Table II. We found the same effects related to the influence of $p$ and $f$. In particular, note that the mass corresponding to the broader resonance peak did not change significantly with $p$. One remarkable point observed with the increase with $p$ is the tendency for the resonances corresponding to tiny peaks to accumulate near the broader peak.

{\tiny
\begin{table}[htbp]
        \begin{tabular}
        {|l |c | c|r|}\hline  & $f=0.5$ & $f=1.1$ & $f=2$  \\
        \hline $p=3$
 & absent & 0.646 &  4.63126; 6.593 \\
        \hline $p=5$ & absent & 0.298164; 0.640 &  5.52762; 6.552 \\
        \hline $p=7$
 & 0.0432 & 0.623; 0.40240& 5.591930; 6.318; 6.872\\
        \hline
        \end{tabular}
               \caption{\it First resonance peaks, even parity modes, left chirality. The table shows the corresponding values of $m^2$.}
   \end{table}}

Now we resume our findings for the resonance spectrum for both chiralities. First of all note that from Eqs. (\ref{fqm}) the spectrum of the right and left chiralities are related. Indeed, the spectrum starts with zero mode with right chirality and even parity. The two resonances with same $m^2$ that succeed in the spectrum are the first odd parity left chiral mode and the first even parity right chiral mode. Next we also have two resonances with same $m^2$: the first even parity left chiral mode and the first odd parity right chiral mode. We successively have in the spectrum even and odd parity wavefunctions for left and right chiral modes with same values of $m^2$. One can easily check the formation of Dirac fermions after studying the odd parity wavefunctions. This needs the changing the normalization procedure in a known procedure \cite{Liu2} (for an analysis with models with two scalar fields, see Ref. \cite{ca}). We checked the correspondence of the spectrum for some values for resonances with odd parity wavefunctions, but we will not pursue this subject in detail, since the numerical procedure was tested with confidence and the formation of the Dirac fermions are guaranteed by the supersymmetric quantum mechanics structure. Also the study of even parity wavefunctions for massive fermions is interesting since we can compare our findings with the results for massive gravitons in the first part of this paper.

\begin{figure}
\includegraphics[angle=-90,width=7.5cm]{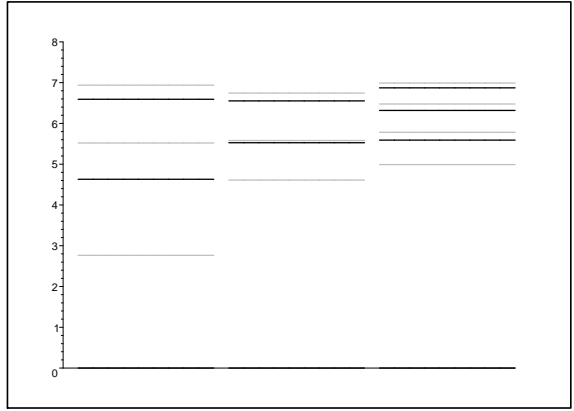}
    \caption{\label{spectrum} Spectrum for $p=3$ (left),  $p=5$ (middle) and $p=7$ (right) with $f=2$ fixed, constructed with even parity modes. Black lines for left chiral massive modes. Grey lines are for right chiral ones.}
\end{figure}

The spectra for $p=3,5,7$ and $f=2$ are depicted in Fig. \ref{spectrum}. There one can see how the increasing in $p$ increases the number of resonances. Note also that the resonances appear coupled in pairs for left and right chiralities (here we consider only even parity wavefunctions). We also see that an increasing of $p$ increases the mass of the first resonance, generally being a very thin line in the spectrum. The fact that the lines for left and right chiralities alternate each other in the spectrum is very important for capture the thinner lines in the numerical procedure.

\section{Conclusions}

In this work we have studied localization of fermions and gravity in a model of deformed branes. The parameter $p$ of the model controls important features as the brane thickness and energy distribution along the extra dimension. We found that the increasing of $p$ cause a splitting in the Ricci scalar, evidencing the appearance of an internal structure, as can be found also from the energy density analysis (see \cite{adalto}). We investigate metric perturbations as decoupled from the scalar ones in the transverse-traceless gauge. From the asymptotic behavior of the Shr\"odinger-like equation it can be proved that gravity is localized for all parameters $p$, and tachyonic gravity modes are absent. Resonances in the gravity sector are observed for $p=1$ and $3$ as peaks in the $|H_{\mu\nu}(0)|^2$ distribution. However, larger values of $p$ lead to produce broader peaks that cannot technically characterize themselves as resonances. This means that thinner branes are more effective for trapping gravity. Next we studied the presence of fermionic zero-modes after introducing a Yukawa coupling between the 5-dimensional spinor and the scalar field, depending on a parameter $f$. After a chiral decomposition we obtained a Shr\"odinger-like equation also for fermionic scalar functions that characterize localization of left or right chiral modes and investigated its normalizability. We found that zero-mode exist only for left chirality and that tachyonic fermionic modes are absent. The investigation of fermionic resonances is a subject possible here only numerically. However, from the qualitative character of the Shr\"odinger potentials we were able to observe that for large values of $f$ the behavior of localization of fermions is favored for larger values of $p$. This is contrary to the observed for gravity, where the localization of KK gravity modes are favored for small values of $p$. The analysis of small values of $f$ could lead to similar behavior for gravitons and fermions. However, the parameter $f$ cannot be reduced below a threshold where the presence of fermionic zero-modes are forbidden. For $f$ above this threshold the effective action is finite. The presence of fermionic resonances was investigated for even parity wavefunctions both for left and right chiralities. We analyzed how the presence of the Yukawa coupling influenced our findings, with larger values of $f$ and $p$ favouring the increasing of resonances. Some resonances are extremely thin and more difficult to be found. The quantum mechanical supersymmetric character of the Shr\"odinger-like potentials for left and right chiral fermions guarantee that Dirac fermions are realized in the model and that the spectral lines for left and right fermions with same parity must alternate in the diagram. This is very helpful in the numerical process for finding resonances, serving as a guide for reducing the numerical step when a resonance line is not captured. The lightest modes in the spectrum correspond to the thinnest lines, showing that those modes couple strongly with the brane in comparison to the KK modes with higher masses. This is expected as lightest modes have lower energy to escape from the brane Shr\"odinger potential. As the inverse of the peak width to half maximum is proportional to the lifetime of the resonance \cite{grs}, thinner peaks may correspond to particles in nature with sufficiently large lifetimes to be important in phenomenology.

The authors would like to thank FUNCAP and CNPq and CNPq/MCT/CT-Infra (brazilian agencies) for financial support.

\end{document}